\documentclass[aps,prx,10pt,longbibliography,superscriptaddress,twocolumn,showpacs,oneside]{revtex4-2}

\usepackage[utf8]{inputenc}
\usepackage[american,british]{babel}
\usepackage[T1]{fontenc}
\usepackage{lmodern}
\usepackage{xcolor}
\usepackage{booktabs}

\usepackage{wrapfig}
\usepackage{graphicx}
\usepackage[caption=false]{subfig}

\usepackage{amsmath,amssymb,amsthm,mathrsfs,mathtools,amsfonts,physics,bm}
\usepackage{mathdots}
\usepackage{dsfont}
\usepackage[version=4]{mhchem}

\usepackage[most]{tcolorbox}
\newtcolorbox{textframe}[2][]{sharp corners, colback=white, boxrule=1pt}

\usepackage[colorlinks=true, linktocpage=true]{hyperref} 
\hypersetup{allcolors = {black}}

\usepackage{comment}
\usepackage[normalem]{ulem}  

\begin{document}
\title{Quantum thermal state preparation for near-term quantum processors}

\author{Jerome Lloyd}
\affiliation{Department of Theoretical Physics, University of Geneva, Geneva, Switzerland}

\author{Dmitry A. Abanin}
\affiliation{Google Research, Brandschenkestrasse 150, 8002 Zürich, Switzerland}
\affiliation{Department of Physics, Princeton University, Princeton NJ 08544, USA}
\affiliation{École Polytechnique Fédérale de Lausanne (EPFL), 1015 Lausanne, Switzerland}


\begin{abstract}

Preparation of quantum thermal states of many-body systems is a key computational challenge for quantum processors, with applications in physics, chemistry, and classical optimization. We provide a simple and efficient algorithm for thermal state preparation, combining engineered bath resetting and modulated system-bath coupling to derive a quantum channel approximately satisfying quantum detailed balance relations. We show that the fixed point $\hat\sigma$ of the channel approximates the Gibbs state as $\|\hat\sigma -\hat\sigma_\beta\|\sim \theta^2$, where $\theta$ is the system-bath coupling and $\hat\sigma_\beta \propto e^{-\beta \hat H_S}$. We provide extensive numerics, for the example of the 2D Quantum Ising model, confirming that the protocol successfully prepares the thermal state throughout the finite-temperature phase diagram, including near the quantum phase transition. Simulations for free-fermion systems provide further evidence for the accuracy of the protocol for large system sizes in the weak-coupling limit. Our algorithm provides a path to efficient quantum simulation of quantum-correlated states at finite temperature with current and near-term quantum processors. 

\end{abstract}

\maketitle

\section{Introduction}

In experimental quantum systems, one frequently observes that the system tends naturally toward a near-equilibrium distribution, so long as heat can be efficiently dissipated into the wider environment. For electronic degrees of freedom in solid state systems, the atomic lattice acts as an effective heat bath, while in atomic and molecular setups, energy typically dissipates into electromagnetic field modes. In contrast, the task of sampling on a quantum computer from the equilibrium (Gibbs) distribution with inverse temperature $\beta$,
\begin{equation}\label{eq:Gibbs}
    \hat\sigma_\beta \propto e^{-\beta \hat H_S},
\end{equation}
is a formidable algorithmic challenge with a wide range of applications \cite{dalzell2025quantum, bergamaschi2024quantum, rajakumar2024gibbs, lemieux2021resource, cao2019quantum, brandao2016quantum}. Given the efficiency of cooling in real-world quantum systems, where does the difficulty arise in designing practical cooling algorithms for use in quantum simulators and quantum processors? 

In particular, two main difficulties have frustrated efforts to design efficient quantum algorithms mimicking cooling processes in nature. Firstly, the basic assumption in statistical mechanics is that the reservoir (or `bath') to which the system is coupled to is macroscopic, with $N_B \gg N_S$, where $N_S$ ($N_B$) is the number of degrees of freedom in the system (bath). Under this assumption, the bath density of states approaches a continuous function, which allows for irreversible decay of the system towards a steady state set by the coupling with the bath \cite{cohen2024photons, breuer2002theory}. A system coupled to a finite bath, on the other hand, will exhibit recurrences in the late-time limit, and does not tend toward a steady state. With present-day quantum simulators of modest sizes, and often restricted geometries, it is not immediately clear how to get around this issue. 

\begin{figure}
    \centering
    \includegraphics[angle=270, width=0.95\linewidth]{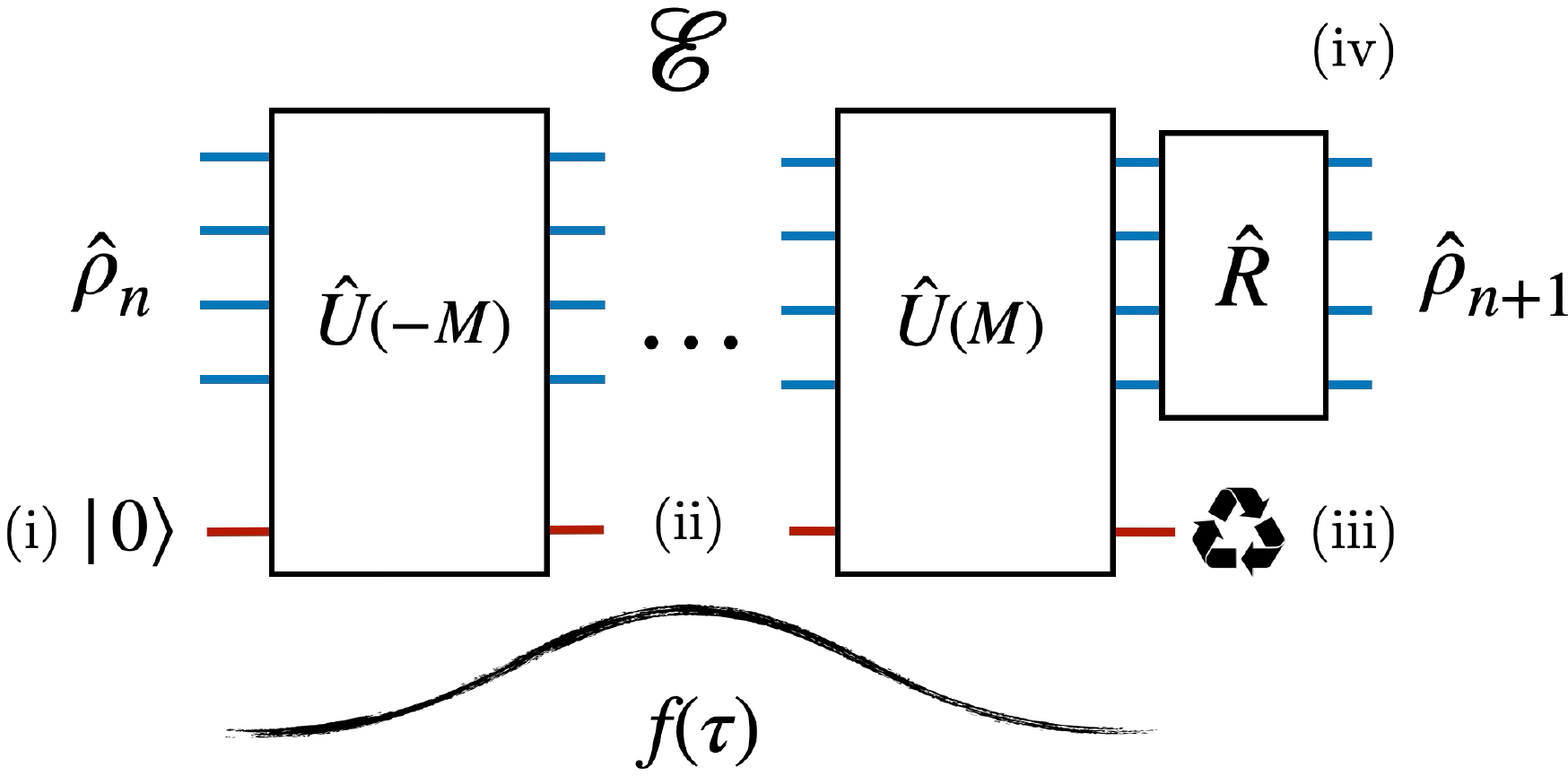}
    \caption{\emph{Protocol for quantum thermal state preparation.} The system density matrix evolves under the repeated map $\hat\rho_{n+1} = \mathcal{E}(\hat\rho_n)$, consisting of three stages: (i) initializing a bath of auxiliary qubits in the $\ket{0}$ state, (ii) joint unitary evolution and (iii) reset of the bath qubits. The unitary stage consists of evolution under a time-dependent unitary $\hat U(\tau)$, defined in Eq.~(\ref{eq:unitary_step}), with $-M \leq \tau \leq M$, $M \propto\sqrt{\beta}$. 
    The coupling to the bath is smoothly switched on with an interaction strength modulated by the filter function $f(\tau)$, to ensure the resulting map approximately satisfies detailed balance. A  randomization step $\hat R$ (iv) grants an additional suppression of unwanted system coherences. Our protocol is designed with near-term analog and digital processors in mind.}
    \label{fig:schematic}
\end{figure}

Secondly, cooling processes are inherently limited by the quantum energy-time uncertainty relation, $\Delta\epsilon \Delta t \gtrsim \hbar$. In order to accurately resolve the energy levels of the system Hamiltonian, one requires times of order of the inverse level spacing, which is typically exponentially large in the system size. This hinders quantum algorithms which depend on quantum phase estimation \cite{kitaev1995quantum} (QPE) routines, such as the quantum Metropolis sampling algorithm introduced in \cite{temme2011quantum}, since the QPE resolution scales inversely proportional to the runtime. We expect many systems to thermalise much faster in practice, at least from the point of view of local observables \cite{lezama2021equilibration}. Exceptions are glassy systems, where the system remains stuck in local minima for extremely long timescales, or low-temperature symmetry breaking phases, where observables probing the symmetry breaking (such as the magnetisation of the Ising model in $d\geq 2$) can take exponentially long times to equilibrate.

Recent works have aimed to surmount these obstacles, with progress in several directions. On the experimental side, recent experiments on a digital Google quantum processor \cite{mi2024stable} have demonstrated preparation of low-energy states of up to 35 qubits, by coupling to a small \emph{resettable} bath of auxiliary qubits. By periodically resetting the bath, the issues associated with recurrences are mitigated, and the system settles into a steady state at late times. A theoretical breakthrough was made in the paper by Chen et al.~\cite{chen2023efficient} (see also \cite{chen2025efficient}), where a Lindblad equation was derived which has the Gibbs state as an exact steady state, but with provably efficient runtime and resource costs. The total Hamiltonian simulation time depends only on the `mixing time' of the Lindbladian, a quantity which essentially measures the worst case time to reach the steady state starting from \emph{any} initial state. Importantly, the mixing time may scale sub-exponentially with system size in many phases, showing that perfect resolution of energy levels is not required for thermal state preparation.  The key insight of \cite{chen2023efficient} is that the conditions of \emph{quantum detailed balance} can be satisfied for a class of efficiently implementable Lindbladians. We refer to \cite{chen2023quantum, chen2021fast, shtanko2021preparing, gilyen2024quantum, ding2025efficient, guo2025designing, hahn2025provably, ding2025end} for related works, as well as \cite{lin2025dissipative} for a pedagogical overview. Unfortunately, the proposed algorithmic techniques are rather high-level and likely out of reach for simulators in the near future (although for very recent progress in this area see \cite{hahn2025efficient}). The mixing time entering the bounds in \cite{chen2023efficient} is also a quantity which is notoriously difficult to bound even in simple cases \cite{levin2017markov}, and is also expected to grow polynomially or exponentially with the system size in most low-temperature quantum phases. Recent efforts have led to impressive new results in this area, in particular proving efficient mixing in certain weakly-interacting or high temperature systems \cite{bardet2023rapid, rouze2025efficient, rouze2024optimal, bakshi2024high, tong2024fast, vsmid2025polynomial}. 

With near-term applications in mind, the current authors put forward a simple algorithm \cite{qpcooling} aimed at efficiently preparing low-energy states of many-body Hamiltonians. The algorithm combined two main features: a small resettable bath as in the experiment \cite{mi2024stable}, and a coupling between system and bath qubits which is \emph{modulated} in time, again with the aim of achieving detailed balance. The algorithm was shown to be effective at preparing ground states of several quantum systems, supported by heuristic arguments for cooling via quasiparticle processes at low temperatures. Other works based on similar principles have helped establish dissipative ground state preparation as a practical and efficient method for future applications \cite{kraus2008preparation, verstraete2009quantum, roy2020measurement, metcalf2020engineered, raghunandan2020initialization, matthies2024programmable, kishony2025gauged, ding2024single, zhan2025rapid, molpeceres2025quantum, stadler2025demonstration}.

In this work, we generalise the modulated coupling protocol of Ref.~\cite{qpcooling} (also dubbed `quasiparticle cooling algorithm') to efficiently prepare quantum thermal states, applicable to present-day digital and analog simulator platforms. There are three essential ingredients to our algorithm, diplayed schematically in Fig.~\ref{fig:schematic}, all of which are accessible on today's hardware: (i) unitary Hamiltonian evolution, $U = e^{-i t \hat H_S}$, implemented either ``exactly'' or via a suitable Trotterisation of the evolution, (ii) fast reset of auxiliary qubits, i.e.~reinitialization to the empty state $\ket{0}$, (iii) time-dependent coupling between the system and auxiliary qubits. The coupling needs only to be of the form $\hat V(t) = \theta f(t) \hat A_i\hat R_j$, where $\hat A_i$ ($\hat R_j$) is a local operator acting on the system (bath), $\theta$ is the overall system-bath coupling strength, and $f(t)$ is a time-dependent modulation. We do not require the operators entering the coupling to vary in time. 

Compared to Ref.~\cite{qpcooling}, our algorithm includes an additional \emph{randomization} step. The extra randomization step involves Hamiltonian evolution for a short random time, and does not change the overall resource-cost scalings of the algorithm. We argue that this step is essential for accurately preparing thermal states, as it suppresses off-diagonal elements in the Hamiltonian eigenbasis. 

For weak system-bath coupling $\theta$, we show, through a combination of perturbative and numerical methods, that the modulated coupling protocol prepares the Gibbs state to an accuracy $\|\hat\sigma_\beta -\hat\sigma\|_1 \sim \theta^2$ in the system-bath coupling, where $\hat\sigma$ is the steady state of the cooling protocol. Thus the errors in both populations and coherences (off-diagonal elements) in the eigenbasis of $\hat H_S$ scale to zero in the limit of small system-bath coupling. Our results hold for any digital or analog platform capable of implementing the simple steps displayed in Fig.~\ref{fig:schematic}.

We numerically demonstrate the efficiency of our protocol by preparing thermal states of the two-dimensional quantum Ising model, for a range of temperatures and interaction strengths, and lattice sizes of up to 16 spins. We benchmark the steady state by comparing several observables against exact values for the thermal state, finding close agreement in all cases. Of particular note is the system heat capacity, which displays clear signatures of the thermodynamic phase transition near the critical values of $\beta$ and $J$ (the ferromagnetic coupling), suggesting that our protocol can be used to prepare quantum-critical thermal states without additional complications. The accuracy of our results increases as the system-bath coupling is decreased, verifying the predicted perturbative scaling. 

We also study thermal state preparation in the example of a free-fermion chain (equivalently, the 1d quantum Ising model). We perform large scale simulations for systems of several hundred sites, assessing the accuracy of the protocol with the system bath coupling and system size. For this model, we demonstate an upper bound on the errors scaling as $\|\hat\sigma-\hat \sigma_\beta\|_1 \leq \mathcal{O}(\theta^2 \sqrt{N_B})$, and and error on local operators such as the energy density $E/N_S$ scaling as $\mathcal{O}(\theta^2)$, independent of system size.

The rest of the paper is organised as follows. In Section \ref{sec:protocol}, we define our modulated coupling protocol for efficiently preparing quantum thermal states. This section can be read as a `recipe' for future digital and analog experiments. In Section \ref{sec:steadystates} we show that the protocol steady state approximates the target Gibbs state, with small errors due to finite system-bath coupling: Subsection \ref{sec:expansion} derives the interaction-picture protocol under the assumption of weak-coupling. Subsection \ref{sec:QDB} explains how the absence of the `coherent term' presented in \cite{chen2023efficient} leads to small violations of quantum detailed balance. Our discussion closely follows that found in \cite{chen2023efficient}. Finally, Subsection \ref{sec:thermal} shows that, after returning to the Schrödinger picture, the violations of detailed balance lead only to small errors on the order of $\mathcal{O}(\theta^2)$ in the system-bath coupling. Turning to numerical results, we present in Section \ref{sec:singlespin} results for single spin cooling, demonstrating the correctness of our perturbative formulas, and in Section \ref{sec:ising} our results for cooling in the 2D Quantum Ising model. We verify the perturbative scaling and show that our protocol accurately prepares the thermal state throughout the finite-temperature phase diagram. In Section \ref{sec:discussion} we provide a short discussion, comparing our protocol with other recently proposed exact and approximate Gibbs samplers.  Finally, in Section \ref{sec:conclusion} we present conclusions and directions for future work. We present our numerical results for free-fermion cooling in Appendix \ref{app:fermions}.

\section{Modulated coupling protocol}\label{sec:protocol}

In this section, we describe our modulated coupling protocol for preparing quantum thermal states. The idea of modulating the coupling to bias cooling transitions was suggested in \cite{qpcooling}, while the idea of engineering a resettable qubit bath goes back to \cite{terhal2000problem}. Throughout this section we use upper case letters with `hats' for operators e.g.~$\hat U$, with the exception of density operators which are denoted either~$\hat \rho$ or $\hat \sigma$. We use $\hat X$, $\hat Y$ and $\hat Z$ to denote the single-site Pauli matrices. Superoperators are denoted with calligraphic notation, e.g.~$\mathcal{E}$. 

We consider $N_S$ system qubits coupled to $N_B$ bath qubits (`auxiliaries'). The bath qubits are prepared initially in the product state $\hat \phi = \ket{0}\bra{0}^{\otimes N_B}$. For the initial state of the system, we take the maximally mixed (infinite temperature) state, $\hat \rho_0 \propto \hat I$. Experimentally, initial states of the system can be uniformly sampled from any complete basis set; in our numerics in section \ref{sec:ising}, we uniformly sample from the computational basis, which is a standard choice experimentally. 

In each step of the cooling protocol, the system and bath are evolved under a joint unitary $\hat Q$, specified below, and then the bath qubits are reset to the initial state $\hat\phi$. The effect of coupling and reset can be expressed as a quantum channel acting on the system density matrix $\hat\rho$,
\begin{equation}\label{eq:resetchannel}
    \mathcal{E}(\hat\rho) = \text{tr}_B [\hat Q(\hat\rho\otimes\hat\phi)\hat Q^\dagger].
\end{equation}
We refer to one application of the above channel as the ``reset cycle''. The system density matrix then evolves as $\hat \rho_{n+1} = \mathcal{E}(\hat\rho_{n})$. The system converges to a fixed point of the channel, defined by $\mathcal{E}(\hat\sigma) = \hat \sigma$, in the limit of a large number of cycles. We assume below that the fixed point is unique, which will be the case for generic systems in the absence of symmetries of the channel $\mathcal{E}$. Note, however, that a system may have long-lived metastable states leading to long mixing times; we discuss this case further below.

So far the specified channel is completely general. We now explain how to choose the unitary $\hat Q$ to prepare quantum thermal states. We focus on the setup for digital platforms, while the setup for analog platforms requires minor modifications which we comment on at the end of this section. We decompose $\hat Q$ into a total of $2M+1$ unitary layers, followed by a `randomisation' unitary $\hat R$: 
\begin{equation}
    \hat Q = \hat R\ \hat U({M})\ldots \hat U({0})\ldots \hat U({-M}).
\end{equation}
Here, $-M \leq \tau\leq M$ is an integer time label. Each unitary $\hat U(\tau)$ is in turn reduced into system (S), bath (B), and coupling components (SB),
\begin{equation}\label{eq:unitary_step}
\hat U(\tau) = \hat U_{SB}(\tau)\hat U_B \hat U_S \equiv e^{-i\delta\theta \hat V({\tau})}e^{-i\delta\hat H_B}e^{-i\delta \hat H'_S},\end{equation}
where $\theta$ is a small dimensionless coupling parameter (the system-bath coupling) and $\delta$ the Trotter angle, also assumed to be small 
in order to avoid errors from Floquet heating effects. Note that the time-dependence of $\hat U(\tau)$ enters only via the time-dependent coupling $\hat V(\tau)$, as will be specified further below.

The Hamiltonian $\hat H'_S$ can be viewed as an effective prethermal Hamiltonian for the system evolution: typically we want to implement $U_S =e^{-i\delta \hat H_S}$ but instead implement $\hat U_S = e^{-i\delta \hat H_1}\ldots e^{-i\delta \hat H_n}$, Trotterising the evolution into $n$ non-commuting gate layers. The prethermal Hamiltonian $\hat H'_S$ then differs from the target system Hamiltonian $\hat H_S$ by a correction which scales as $\mathcal{O}(\delta)$ and which is the sum of (quasi)-local terms \cite{yin2023prethermalization}. In driven systems, although the energy is not conserved, the prethermal Hamiltonian operator $\hat H'_S$ is typically conserved for times which are exponentially long in the driving frequency $\omega_\delta = 2\pi/\delta$ (relative to the natural energy scales of the problem) \cite{abanin2017effective, yin2023prethermalization}. Replacing $\hat U_S$ with higher-order Trotter decompositions \cite{childs2021theory} is expected to effectively reduce the errors due to finite $\delta$, although high-order expansions may not be practical on current devices. Here, we do not aim to provide a detailed analysis of the errors due to finite Trotter angles, and will mainly consider the Hamiltonian cooling limit obtained as $\delta\to 0$, $M\to \infty$, and $\hat H'_S \to \hat H_S$, with the ratio $T\equiv M\delta$ kept constant. In the rest of the paper we write $\hat U_S = e^{-i\delta \hat H_S}$ for the system unitary, with the above limit assumed. We refer to parameter $T$, which  measures the overall depth of the unitary evolution before the bath reset takes place, as the ``reset time''. 

The randomisation unitary $\hat R$ consists of a random-depth unitary acting only on the system, 
\begin{equation}\label{eq:random_unitary}
    \hat R = e^{-i\delta \hat H'_S M_R} = \hat U_S^{M_R},
\end{equation}
where $M_R \in [0,\infty)$ is a random integer typically on the same order as $M$, sampled at the start each new reset cycle. For concreteness we will consider the case where the probability $p(M_R)$ follows the exponential distribution
\begin{equation}\label{eq:exp_distribution}
    p(M_R) \propto e^{-M_R/\lambda M}.
\end{equation}
$\lambda$ parametrizes the distribution of $M_R$ relative to $M$, and will be referred to as the ``randomization parameter''. The reason for the extra randomisation step will be explained in detail in Section \ref{sec:thermal}: essentially, it acts as a further dephasing map on the density matrix coherences. In our results below, we will make use of both the protocol with the extra randomization step (randomized protocol) and without (unrandomized protocol), in which case we set $\lambda=0$ in Eq.~(\ref{eq:exp_distribution}). Similar randomized evolution times have have been considered in Refs.~\cite{shtanko2021preparing, molpeceres2025quantum}.

For the bath Hamiltonian, a sufficient choice is to take non-interacting qubits according to 
\begin{equation}
    \hat H_B = -\frac{h}{2}\sum_{\mu=1}^{N_B} \hat Z_\mu,
\end{equation}
where parameter $h$ enters as the single bath energy scale. It is also possible (and may be advantageous in some settings \cite{shtanko2021preparing, molpeceres2025quantum}) that each bath spin is subject to an independent field $h_\mu$, as long as the system-bath couplings are adjusted accordingly (see Eq.~(\ref{eq:filterfunction})). 

The $\tau$-dependent coupling operator $\hat V({\tau})$ is specified by 
\begin{equation}\label{eq:Vdef}
   \hat V({\tau}) = f(\tau)\sum_{\mu = 1}^{N_B}  \ \hat A_{S,\mu} \otimes  \hat Y_{B,\mu},
\end{equation}
where $f(\tau)$ is a modulating function which will be referred to as the `filter function', $\hat Y_{B,\mu}$ is the Pauli-$Y$ matrix acting on the $\mu$-th auxiliary qubit ($\hat X_{B,\mu}$ can also be used), and $\hat A_{S,\mu}$ is a local system operator associated with the $\mu$-th auxiliary coupling, which is assumed to be Hermitian. Below, we will simply write $\hat A_\mu \equiv \hat A_{S,\mu}$, with the understanding that $\hat A_\mu$ acts on the system. The precise choice of $\hat A_\mu$ does not affect arguments regarding the form of the steady state (after making the Hermiticity assumption), but a good choice of the operators is important for controlling the ergodicity of the dynamics, as well as ensuring fast approach to the steady state. For preparing low-temperature states, it is beneficial to choose $\hat A_\mu$ to target quasiparticle-like excitations of the system, while still remaining local \cite{qpcooling}. Frequently, a single Pauli operator suffices, e.g.~$\hat A_\mu = \hat Y_{i_\mu}$.   

The choice of the filter function is the critical element in designing the protocol such that the steady state approaches the Gibbs state. The class of allowed functions is restricted according to the principle of quantum detailed balance \cite{ding2025efficient}, as will be explained in Section \ref{sec:QDB}. This still leaves some flexiblity, and a convenient choice is to take a Gaussian filter, 
\begin{equation}\label{eq:filterfunction}
    f(\tau) \propto \exp\Big(-\frac{a^2\delta^2\tau^2}{2}\Big)
\end{equation}
with a normalisation $\delta \sum_{\tau = -M}^{M} |f(\tau)| = 1$, and width set by 
\begin{equation}\label{eq:adef}
    a = \sqrt{\frac{4 h}{\beta}}.
\end{equation}
While at this level $\beta$ enters as a free parameter of the protocol, the association with the inverse temperature of the Gibbs state is not accidental. The bath splitting $h$ enters to fix the overall width of the filter function. In order to avoid truncation effects from the finite reset time, we require $T \gg \sqrt{\beta/4h}$. Thus low temperatures (large $\beta$) require longer periods of unitary evolution. In our numerics, we have found $Ta\lesssim \mathcal{O}(1)$ sufficient for effective cooling, due to the fast decay of the Gaussian tails. The bath parameter, $h$, should also be chosen judiciously, as it fixes the resonance centre for the cooling processes (see Eq.~(\ref{eq:filter_frequency})) and thus affects the rate at which the steady state is approached. Normally we should choose $h \gtrsim \Delta$, where $\Delta$ is the many-body energy gap, to ensure resonant removal of low-energy excitations \cite{qpcooling} --- we have found this choice efficient in our numerical examples. In applications the system gap is usually not known precisely. However, simple estimates are often available and the parameter $h$ may be varied experimentally to find approximately optimal values of $h$.

Above we defined the cooling protocol for use on digital quantum simulators. The protocol can be defined similarly for analog machines. In this case we replace the unitary evolution by 
\begin{equation}
    \hat Q = \hat R\ \mathcal{T} e^{-i\int_{-T}^{T}dt\ (\hat H_S+\hat H_B+ \theta \hat V(t))},
\end{equation}
where $\hat V(t)$ is defined by Eq.~(\ref{eq:Vdef}) with the replacement $\tau\delta  \to t$ and $\hat R = e^{-iT_R\hat H_S}$. In the next section, we will work exclusively in the $\delta\to0$ limit to simplify our derivations, so the results hold for both analog protocols and digital protocols (in the limit of small Trotter step). We will abuse notation slightly by writing e.g.~$f(t)$, $\hat A(t)$ for continuous-time versions of functions where we replace $\tau \delta \to t$, but the meaning should be clear.

Experimentally, implementing our protocol on analog simulators requires the capability for reset of the bath qubits, which is available e.g.~in the recently developed digital-analog superconducting qubit platform~\cite{AndersenNature2025Analog}, or trapped ion architectures \cite{barreiro2011open}. Digital processors allow a rapid reset on timescales comparable to two-qubit gates operation times, see e.g.~\cite{mcewen2021removing, miao2022overcoming, mi2024stable}.

In summary, our protocol requires: (i) time evolution under the system unitary $U_S = e^{-it \hat H_S}$, for a time $T = \mathcal{O}(\sqrt{\beta/h})$ (ii) a small ($N_B \geq \mathcal{O}(1)$) bath of resettable qubits (in practice $N_B \propto N_S$ is likely needed to efficiently compete with device noise) and (iii) weak local coupling between system and bath, with time-dependent interaction strength set by the filter function, Eq.~(\ref{eq:filterfunction}). The parameters of our protocol are displayed in Table \ref{table} for convenience. 

\begin{table}[t]\label{table}
\centering
\begin{tabular}{c@{\hspace{0.6em}}|@{\hspace{0.6em}}l}
\toprule
Parameter & Meaning \\
\midrule
$\hat A_\mu$ & cooling operators ($\mu$-th coupling) \\ 
$N_B$ & number of bath spins \\
$N_S$ & number of system spins \\
$\beta$ & target inverse temperature \\
$\theta$ & system–bath coupling strength \\
$h$ & bath energy scale \\
$\delta$ & Trotter angle \\
$T $ & reset time \\
$\lambda$ & randomization parameter \\
$a$ & filter function width, Eq.~(\ref{eq:adef})\\
\bottomrule
\end{tabular}
\caption{Parameters of the protocol and their symbols used throughout the text (note not all parameters in the table above are independent).}
\end{table}

\section{Approximately thermal steady states}\label{sec:steadystates}

To which extent do the steady states of the cooling protocol $\mathcal{E}$ approximate the target thermal state, $\hat \sigma_\beta$? To answer this question, we will first show that the modulated coupling protocol leads to small violations of `detailed balance' in the interaction frame defined with respect to the system Hamiltonian $\hat H_S$. These violations cause the steady state to deviate from the exact Gibbs state. We then calculate the errors introduced in the steady state by means of a perturbation theory around the Gibbs state, and show that the errors generically scale as $\|\hat \sigma-\hat\sigma_\beta\|_1 \sim \theta^2$, in terms of the system bath coupling. 

\subsection{Expansion in weak coupling}\label{sec:expansion}

Intuitively, we expect errors in the thermal state to arise due to finite coupling between the system and the bath.  We therefore consider the weak-coupling limit $\theta^2 \ll 1$ where we may 
expand the reset channel, Eq.~(\ref{eq:resetchannel}), in powers of the system-bath coupling. Due to our choice of normalisation for the filter function, $\theta$ is dimensionless and the relevant small parameter to control the weak-coupling limit.    

The interaction picture is defined with respect to the free system-bath evolution by
\begin{equation}\label{eq:bare_interaction_picture}
   \hat O_I(\tau) = \hat U_0^{-\tau} \hat O(\tau)\hat U_0^\tau, \hspace{.5cm} \hat U_0 = \hat U_B\hat U_S,
\end{equation}
such that 
\begin{gather}
    \hat Q_I = \hat U_S^{-M_R} \hat U_0^{-M} \hat Q \hat U_0^{-(M+1)} \nonumber\\ = \hat {U}_{SB,I}(M)\ldots\hat {U}_{SB,I}(0)\ldots \hat{U}_{SB,I}(-M).
\end{gather} 
We note that this unconventional definition of the interaction picture is due to our choice of symmetrising the reset protocol with respect to $-M \leq \tau \leq M$, and is purely convention. The interaction picture allows us to rewrite Eq.~(\ref{eq:resetchannel}) as the concatenation of channels
\begin{gather}
    \mathcal{E}(\hat\rho) = 
    \mathcal{S}^{M+M_R} \circ{\mathcal{E}}_I\circ\mathcal{S}^{M+1}(\hat\rho),\label{eq:Schrödingermap}
\end{gather}
where 
\begin{equation}
    \mathcal{S}(\hat\rho) = \hat U_S(\hat\rho)\hat U_S^{-1},
\end{equation}
\begin{equation}\label{eq:interactionmapdef}
   {\mathcal{E}}_I(\hat \rho) = \text{tr}_B\big[ \hat{Q}_I(\hat\rho\otimes \hat\phi)\hat{Q}^{\dagger}_I\big].
\end{equation}
We refer to ${\mathcal{E}}_I$ as the interaction-picture map and $\mathcal{E}$ as the Schr\"odinger picture map. We will not distinguish between states in the interaction-picture vs.~Schr\"odinger picture, since the action of Eq.~(\ref{eq:interactionmapdef}) is always defined on its input.

Next, under the weak-coupling assumption, we expand $\mathcal{E}_I$ to second order in $\theta$ (we leave the details to Appendix \ref{app:expansion}) to arrive at the interaction-picture map in Lindblad form:

\begin{equation}\label{eq:lindblad}
    \frac{\mathcal{E}_I(\hat\rho)-\hat{\rho}}{\theta^2} = -i[\hat G^{\text{LS}}, \hat\rho]- \{\hat K,\hat\rho\}+\sum_\mu \hat L_\mu \hat\rho\hat L_\mu^\dagger.
\end{equation}
Here,
\begin{equation}\label{eq:jumpdef}
    \hat L_\mu = \delta \sum_{\tau=-M}^{M}  f(\tau) e^{i\delta h\tau}\hat A_{\mu,I}(\tau), 
\end{equation}
is the jump operator associated to interactions with the $\mu$-th auxiliary, $\hat K = \frac{1}{2}\sum_\mu \hat L^\dagger_\mu\hat L_\mu$, and $\hat G^{\rm LS} = \sum_\mu \hat G^{\rm LS}_\mu$
is a Hamiltonian correction traditionally referred to as the ``Lamb shift'', defined in Appendix \ref{app:expansion}, Eq.~(\ref{eq:GLSderivation}).  

Since $T\gg a^{-1}$, the limits on the Gaussian sums may be extended to infinity with marginal error. Using also the assumption of small $\delta$ (we require $\delta a \ll 1)$, we make the replacement $\tau\delta \to t$ and replace the summation formulae with continuous-time integrals according to
\begin{equation}\label{eq:jump_continuous}
    \hat L_\mu = \int_{-\infty}^\infty dt\ f(t)e^{iht} \hat  A_{\mu,I}(t),
\end{equation}
and similarly for the Lamb shift. The error incurred in this replacement is $\text{max}[\mathcal{O}(e^{-T^2a^2}/aT), \mathcal{O}(e^{-\pi^2/2\delta^2a^2})]$ as shown in Appendix \ref{app:fourier}. The final form of the interaction-picture map is the Lindblad equation specified in Eq.~(\ref{eq:lindblad}), with the operators replaced with their continuous time versions. 

\subsection{Violations of quantum detailed balance}\label{sec:QDB}

As we will show below, the steady states of the Lindblad equation, Eq.~(\ref{eq:lindblad}), are not exact Gibbs states with respect to the system Hamiltonian $\hat H_S$. The error can be traced to an `incorrect' choice of the Lamb-shift Hamiltonian, $\hat G^{\text{LS}}$. It was shown in \cite{chen2023efficient} that, surprisingly, there exists a different choice of Hamiltonian, $\hat G^{\text{DB}}$, which ensures that the steady state is \emph{exactly} thermal. Below we give a simple proof of this statement. In our case, the Lamb-shift is fixed by the cooling protocol, and it is not clear how to derive a similar protocol where $\hat G^{\text{DB}}$ appears naturally in the interaction picture.  

We first introduce the concept of quantum detailed balance (QDB) \cite{guo2025designing}, following the definition in \cite{temme2011quantum}, which holds for a general quantum channel $\mathcal{E}$. This coincides with the definition of `KMS' detailed balance used in other works, defined at the level of the Lindbladian \cite{chen2023efficient, ding2025efficient}. For a given basis, $\ket{\psi_a}$, the QDB condition requires that $\mathcal{E}$ satisfies the relation 
\begin{equation}\label{eq:qdb}
     \frac{\bra{\psi_a}\mathcal{E}(\ket{\psi_c}\bra{\psi_d})\ket{\psi_b}}{\bra{\psi_d}\mathcal{E}(\ket{\psi_b}\bra{\psi_a})\ket{\psi_c} } = \sqrt{\frac{p_a p_b}{p_c p_d}},
\end{equation}
where $p_a$ are probabilities satisfying $\sum_a p_a = 1$. Then, 
\begin{equation}
    \hat{\pi} = \sum_a p_a \ket{\psi_a}\bra{\psi_a} 
\end{equation}
is a fixed point of the channel $\mathcal{E}$, since 
\begin{gather}
    \bra{\psi_a} \mathcal{E}(\hat\pi)\ket{\psi_b} = \sqrt{p_ap_b} \sum_c \bra{\psi_c }\mathcal{E}(\ket{\psi_b}\bra{\psi_a}) \ket{\psi_c } \nonumber \\ 
    = p_a\delta_{ab},
\end{gather}
where we used the fact that $\mathcal{E}$ is trace-preserving. We say that $\mathcal{E}$ satisfies quantum detailed balance with respect to the state $\hat \pi$.

Our choice of filter function in Eq.~(\ref{eq:filterfunction}) is made in order to approximately satisfy the QDB criteria with respect to the Gibbs state, $\hat \sigma_\beta$. To see this, consider the action of $\hat L_\mu$ between eigenstates of the system Hamiltonian, $\ket{\phi_a}$. For the jump operators defined in Eq.~(\ref{eq:jump_continuous}), 
\begin{gather}
    \bra{\phi_a} \hat L_\mu \ket{\phi_b} 
    =  A_{\mu ab} \bar f_h(\omega_{ab}),\label{eq:jump_transition1}
\end{gather}
where $\omega_{ab} = \epsilon_b-\epsilon_a$ are the transition (Bohr) frequencies, $A_{\mu a b} = \bra{\phi_a}\hat A_\mu \ket{\phi_b}$, and
\begin{equation}\label{eq:filter_frequency}
    \bar f_h(\omega) = \int _{-\infty}^\infty dt\ e^{-i(\omega-h) t} f(t) = e^{-\frac{(\omega-h)^2}{2a^2}}.
\end{equation}
It is easy to check that the function $\bar f_h(\omega)$ satisfies the \emph{classical} detailed balance condition,
\begin{equation}
    \bar f_h(\omega) = e^{\beta\omega/2} \bar f_h(-\omega)
\end{equation}
using our choice of $a$ in Eq.~(\ref{eq:adef}). This fixes the jump operators to obey the relation
\begin{equation}\label{eq:dbjump}
    \bra{\phi_a} \hat L_\mu \ket{\phi_b}  = e^{\beta\omega_{ab}/2} \bra{\phi_a} \hat L^\dagger_\mu \ket{\phi_b}. 
\end{equation}
We then separate the Lindblad equation in (\ref{eq:lindblad}) into two parts, the dissipative part 
\begin{equation}
    \mathcal{W} = \sum_\mu \hat L_\mu \bullet \hat L^\dagger_\mu, 
\end{equation}
and a part 
\begin{equation}
    \mathcal{G} = -i[\hat G,\bullet] +\{\hat K,\bullet\} 
\end{equation}
which depends on a general Hamiltonian term $\hat G$ in place of $\hat G^{\text{LS}}$. From Eq.~(\ref{eq:dbjump}), QDB is satisfied for the dissipative part:
\begin{equation}
    \frac{\bra{\phi_a}\hat L_\mu(\ket{\phi_c}\bra{\phi_d})\hat L^\dagger_\mu\ket{\phi_b}}{\bra{\phi_d}\hat L_\mu(\ket{\phi_b}\bra{\phi_a})\hat L^\dagger_\mu\ket{\phi_c} } = \sqrt{\frac{e^{\beta\epsilon_d/2}e^{\beta\epsilon_c/2}}{{e^{\beta\epsilon_a/2}e^{\beta\epsilon_b/2}}}}
\end{equation}
with $p_a \propto e^{-\beta \epsilon_a}$ the Gibbs weights. On the other hand, $\mathcal{G}$ satisfies QDB only if 
\begin{gather}
    \sqrt{p_b } \bra{\phi_a}\hat J^\dagger \ket{\phi_b} = \sqrt{p_a} \bra{\phi_a}\hat J \ket{\phi_b},
\end{gather}
with the non-Hermitian operator
\begin{equation}
    \hat J = \hat K + i\hat G.
\end{equation}
Solving this equation for $\hat G$, we find 
\begin{equation}\label{eq:trueLamb}
     \hat G^{\text{DB}}_{ab} = -i\tanh \bigg(\frac{\beta\omega_{ab}}{4}\bigg)\hat K_{ab}.
\end{equation}

For this choice of the coherent term, the Lindblad equation 
\begin{equation}\label{eq:DBlindblad}
    {\mathcal{L}}^{\text{DB}} = \mathcal{G}^{\text{DB}}+\mathcal{W}, \hspace{.5cm} \mathcal{G}^{\text{DB}} = -i[\hat G^{\text{DB}},\bullet] +\{\hat K,\bullet\} 
\end{equation}
satisfies QDB and hence $\mathcal{L}^{\text{DB}}(\hat\sigma_\beta) = 0$. Note also that $\hat G^{\text{DB}}$ is only unique up to the addition of a term $\hat \Lambda$ which is diagonal in the eigenbasis of $\hat H_S$, since any such term commutes with the Gibbs state. 

Unfortunately, our Lamb-shift does not have the form of Eq.~(\ref{eq:trueLamb}). We give an explicit expression for the matrix elements of $\hat G^{\text{LS}}$ in Appendix \ref{app:Lamb}, from which it is readily seen that the two differ. We denote the error as 
\begin{equation}
    \Delta \hat G = \hat G^{\text{LS}}-\hat G^{\text{DB}}, 
\end{equation}
and for non-zero $\Delta \hat G$, the QDB conditions are not satisfied. While it is possible that quantum channels may exist which have the Gibbs state as their fixed point despite not satisfying the detailed balance conditions (this statement is at least true for classical Markov chains), this is not true in our case, as we show in the next section.  

\subsection{Approximately thermal steady states }\label{sec:thermal}

Due to the violations of detailed balance, the steady state of the protocol differs from the Gibbs state. We denote the protocol steady state by $\hat\sigma$ and the error by $\hat\zeta \equiv \hat \sigma-\hat\sigma_\beta$. In this section, we will show that the errors scale with the system-bath coupling and at weak coupling the steady state is approximately thermal. This claim is not completely obvious. While corrections to the detailed balance Lindbladian are small, $\mathcal{O}(\theta^2 )$, the thermalisation timescale is long, $\mathcal{O}(1/\theta^2 )$. The sum of many small errors can therefore accumulate in the steady state. Our resolution comes only after returning to the Schrödinger picture: the perturbation generated by $\Delta\hat G$ is off-diagonal in the Hamiltonian eigenbasis, and since off-diagonal elements between widely separated energy levels are rapidly dephased with a rate $\gamma_\omega \sim \omega T$, the small errors cannot accumulate coherently. Our statement can equivalently be interpreted as saying that dropping the `coherent term', $-i[\hat G^{DB},\bullet]$, introduced in Refs.~\cite{chen2023efficient, chen2025efficient} to exactly restore quantum detailed balance, leads only to perturbatively small errors after moving back to the Schr\"odinger picture. 

To perform the perturbative expansion, we move back to the Schrödinger picture via Eq.~(\ref{eq:Schrödingermap}). To simplify final expressions, we consider the small $\delta$ regime, where we can replace $M\delta = T$ and assume $\omega_{ab}\delta\ll1$. 
The Schrödinger map contains the term $\mathcal{S}^{M_R}$ which depends on the randomization step $M_R$. In order to obtain the steady state after repeated application of the random channel, we consider the averaged map. With $p(M_R)$ defined in Eq.~(\ref{eq:exp_distribution}), the average of the unitary channel $\mathcal{S}^{M_R}$ now implements a dephasing channel in the eigenbasis of the Hamiltonian. Defining $\mathcal{D} \equiv \mathbb{E}_{M_R}[\mathcal{S}^{M_R}]$, and setting $T_R = M_R\delta$, we have
 \begin{gather}
     \mathcal{D}(\hat\rho) = \frac{1}{\lambda T} \int_0^\infty dT_R\ e^{-T_R/\lambda T} e^{-i\hat H_ST_R} (\hat\rho) e^{i\hat H_ST_R} \nonumber \\
     = \sum_{ab} \frac{\ket{\phi_a}\bra{\phi_a}\hat\rho\ket{\phi_b}\bra{\phi_b} }{1-i\omega_{ab}\lambda T}\label{eq:averagedR},
 \end{gather}
 with $\lambda$ measuring the `strength' of the randomization step. For $\lambda = 0$ the protocol is unrandomized. 

The Schrödinger map can then be written as
\begin{equation}
    \mathcal{E}(\hat\rho) = \mathcal{D}\circ\mathcal{E}_0 \circ \mathcal{E}_I\circ \mathcal{E}_0(\hat \rho), 
\end{equation}
where $\mathcal{E}_0 = U_S^T(\bullet)U^{\dagger T}_S$ and we write $\mathcal{E}_I$ as
\begin{equation}
    \mathcal{E}_I(\hat\rho) \approx  \Big(1+\theta^2\mathcal{L}\Big)\hat \rho,
\end{equation}
\begin{equation}
    \mathcal{L}(\hat\rho) = -i[\Delta\hat G, \hat\rho] + \mathcal{L}^{\text{DB}}(\hat\rho).
\end{equation}

For $\Delta \hat G = 0$, the steady state is the Gibbs state $\hat \sigma_\beta$ and $\hat \zeta = 0$. Away from this limit, the correction to the Gibbs state is obtained by
\begin{gather}
    \hat\zeta =  \sum_{n=0}^\infty(\mathcal{E}^{n+1}(\hat\sigma_\beta) - \mathcal{E}^n(\hat\sigma_\beta)) = 
   \sum_{n=0}^\infty\mathcal{E}^{n}(\hat \xi), \label{eq:zetadef}
\end{gather}
where $\hat \xi = -i\theta^2 \mathcal{D}\circ\mathcal{E}_0( [\Delta\hat G,\hat\sigma_\beta])$ is the `perturbation' seeded by $\Delta\hat G$. Note that $\hat \xi$ is traceless, has vanishing diagonal in the Hamiltonian eigenbasis, and $\|\hat \xi \|_1 \lesssim \mathcal{O}(\theta^2 N_B)$ \footnote{A rigorous bound on the commutator norm can be obtained using similar methods to \cite{hahn2025provably}.}, with $\| \hat M\|_1$ the trace norm of $\hat M$. A worst-case bound on $\|\hat \zeta\|_1$ can be derived by defining the mixing time $t'_{\rm mix}$ of the channel $\mathcal{E}$ as the smallest $s$ such that
\begin{equation}\label{eq:mixing}
\|\mathcal{E}^n(\hat X)\|_1 \le e^{-n/s}\,\|\hat X\|_1
\qquad 
\end{equation}
holds, for all traceless operators $\hat X$, assuming the steady state is unique. It follows that $\|\hat \zeta\|_1 \lesssim t'_{\rm mix}\|\hat \xi\|_1$. However, since thermalisation occurs perturbatively slowly in $\theta$, the mixing time is expected to scale as $t'_{\rm mix} \sim t_{\rm mix}/\theta^2$, where $t_{\rm mix}$ is an `intrinsic' timescale which does not depend (strongly) on $\theta$. This gives the trivial bound 
\begin{equation}\label{eq:naive_mixing_bound}
    \| \hat \zeta\|_1 \lesssim \mathcal{O}(N_B t_{\rm mix}).
\end{equation}

Below, we argue that errors are perturbatively small in $\theta^2$, in spite of the above bound, which is too loose. This arises due to the following observations: errors which are diagonal in the Hamiltonian eigenbasis relax perturbatively slowly due to the system-bath interaction, but the initial perturbation $\hat\xi$ has no diagonal component. Errors which are off-diagonal relax non-perturbatively, due to dephasing in the Schr\"odinger picture. This leads to the stronger bound $\|\zeta\|_1 \lesssim \mathcal{O}(\theta^2 \tilde g(N_B,N_S) t_{\rm mix})$, where $\tilde g(N_B,N_S)$ is a function introduced below. Although we are not able to determine $\tilde g$ explicitly, in many cases we expect $\tilde g \lesssim  N_B$, i.e.~the error grows at most extensively with the size of the bath, and the dependence on system size only enters through the mixing time.

To see the role of dephasing, we rewrite the series in Eq.~(\ref{eq:zetadef}) using the Dyson series, 
\begin{equation}\label{eq:Dyson}
    \hat \zeta = \mathcal{K}_0(\hat \xi) + \theta^2\mathcal{K} \circ \mathcal{L}\circ \mathcal{K}_0(\hat \xi),
\end{equation}
where $\mathcal{K} = \sum_{n=0}^\infty \mathcal{E}^n$, $\mathcal{K}_0 = \sum_{n=0}^\infty (\mathcal{D}\circ \mathcal{E}_0^2) ^n$ are the interacting and free resolvents respectively. The motivation is to separate the dynamics into `slow' and `fast' subspaces, which approximately govern the mixing of populations and the dephasing of coherences \footnote{We thank Dominik Hahn for useful related discussions. In a previous version of the manuscript, we derived the errors via a direct perturbation theory, but this did not allow us to estimate the populations.}. Importantly, we expect terms entering the sum $\mathcal{K}_0(\hat \xi)$ to decay faster than the perturbative timescale $\sim 1/\theta^2$. 

We define $\hat\zeta^{\rm coh}$ and $\hat\zeta^{\rm pop}$ respectively as the corrections to coherences and populations, i.e. the off-diagonal and diagonal projections of $\hat\zeta$ in the Hamiltonian eigenbasis. Since $\mathcal{K}_0(\hat \xi)$ is completely off-diagonal, it sets the correction to coherences to leading order in $\theta$, $\hat\zeta^{\rm coh} \approx \mathcal{K}_0(\hat \xi)$. (When projecting $\hat \zeta$ onto the off-diagonal subspace, we can replace $\mathcal{K}$ with $\mathcal{K}_0$ to leading order; hence the second term in Eq.~(\ref{eq:Dyson}) is sub-leading.) Meanwhile the correction to populations is given by the second term, $\hat \zeta ^{\rm pop} = \theta^2\mathcal{K} \circ \mathcal{L}\circ \mathcal{K}_0(\hat \xi)$. Note that it is not possible to replace $\mathcal{K}$ with $\mathcal{K}_0$ when calculating the populations, since this formally leads to a divergence. 

We can explicitly calculate the correction to the coherences by summing $\mathcal{K}_0(\hat \xi)$, 
\begin{equation}
    \zeta^{\rm coh}_{a\neq b} = i\theta^2(\Delta \hat G)_{ab}\bigg(\frac{p(\epsilon_b) - p(\epsilon_a)} {1 - e^{2iT\omega_{ab}}-i\omega_{ab}\lambda T}\bigg), \label{eq:corrections}
\end{equation}
where $p(\epsilon) = e^{-\beta \epsilon}/{Z}$ is the Gibbs weight with $Z = \text{Tr}(e^{-\beta \hat H_S})$. Coherences between well-separated energy levels are dephased on a characteristic timescale $t_\omega \sim (\omega T)^{-1}$. On the other hand, coherences between nearby levels dephase slowly, but are suppressed in the initial perturbation by the Gibbs weight factors. The suppression guarantees that the series does not diverge as the spacing $\omega_{ab}$ is taken to zero --- this should be contrasted to the case where $\hat \xi$ is a generic perturbation, for which $\mathcal{K}_0$ diverges with the inverse level spacing. 

Eq.~(\ref{eq:corrections}) highlights the role played by the additional randomization step: for $\lambda = 0$, we find divergences at the special values $\omega_{ab} = m \pi/T$ when the frequency is resonant with the drive frequency (this effect survives for $\delta \to 0$, and is not related to the discretization of the protocol). The randomization is added to wash out this resonant effect. For finite $\lambda$, the size of the coherences in Eq.~(\ref{eq:corrections}) can be bounded. Assuming $\lambda \geq 2$, and $\epsilon_b > \epsilon_b$ for simplicity, we obtain the bound (see Appendix \ref{app:bounds}):
\begin{equation}\label{eq:coherence_bound}
    |\zeta^{\rm coh}_{a\neq b}| \leq C_1  \theta^2N_B e^{-\omega_{ab} ^2/4a^2}\min \bigg[ \frac{\beta}{2T}, \frac{1}{|\omega_{ab}|T}\bigg] p(\epsilon_a),  
\end{equation}
where $C_1 \sim \mathcal{O}(1)$. The above bound shows that coherences are perturbatively small in $\theta$. 

While we are unable to directly translate Eq.~(\ref{eq:coherence_bound}) into a useful bound on the trace norm, we have formally $\|\hat \zeta^{\rm coh}\|_1 \leq  \theta^2 g(N_S,N_B)$, where $g(N_S,N_B)$ is a function depending on the number of system and bath spins (as well as other parameters of the protocol which we suppress). 

Next we consider the population corrections. Since $\mathcal{L}(\hat \zeta^{\rm coh})$ remains in the traceless subspace, the correction can be bounded using Eq.~(\ref{eq:mixing}) by 
\begin{equation}\label{eq:population_error}
    \|\hat \zeta^{\rm pop} \|_1 \leq \theta^2 t'_{\rm mix} \|\mathcal{L}(\hat\zeta^{\rm coh})\|_1 \leq \theta^2 \tilde{g}(N_S,N_B) t_{\rm mix},
\end{equation}
with $\tilde{g}(N_S,N_B) = g(N_S,N_B)\times C_2 N_B$ and we used $\|\mathcal{L}(\hat \zeta^{\rm coh})\|_1 \leq C_2 N_B\|\hat \zeta^{\rm coh}\|_1$. The contribution from populations is naively a factor of $N_B t_{\rm mix}$ larger than the contribution from coherences. Combining the two contributions, and absorbing constants into $\tilde{g}$, we have the scaling
\begin{equation}\label{eq:tracebound}
    \|\hat \zeta \|_1 \lesssim \theta^2 \tilde{g}(N_S,N_B)t_{\rm mix}.
\end{equation}
We expect in many physically relevant cases $\tilde g \lesssim N_B$, which is the scaling one gets from the naive mixing time bound Eq.~(\ref{eq:naive_mixing_bound}). Furthermore, any mechanism which suppresses the size of the coherences, such as our randomization step, is expected to lead to an improved bound in Eq.~(\ref{eq:tracebound}). Any non-trivial scaling with the number of qubits is expected to come from the mixing time. 
In Appendix~\ref{app:fermions}, for a free-fermionic chain (which is expected to mix rapidly, $t_{\rm mix} \lesssim \log N_S$ \cite{vsmid2025rapid}) and a wide range of parameters we numerically demonstrate the scaling $\|\hat \zeta\|_1 \approx \mathcal{O}(\theta^2 \sqrt{N_S})$.

We now comment on the experimentally relevant regimes of low and high temperature. We consider gapped many-body systems with gap $\Delta$, and estimate the size of corrections to the Gibbs state:

\textbf{Low temperature}: In the limit $\Delta \beta \gg 1$, we aim to cool toward the ground state, which has been successfully demonstrated in several recent works  \cite{qpcooling, ding2024single, zhan2025rapid} using variants of the cooling protocol.  The population errors are bounded by Eq.~(\ref{eq:population_error}), which may be large if the mixing time is large. However, since $\hat \xi$ is exponentially suppressed at high energies by the Gibbs weight, we expect the corrections at high energies to be similarly small, unless the relaxation time is very large or $\mathcal{L}$ directly couples low and high energy subspaces. 

Since coherences at large energies are suppressed by the Gibbs weight entering Eq.~(\ref{eq:coherence_bound}), we now focus on the coherence between the ground state and low-lying excited states. In this case we have $|\omega_{0b}|  \geq \Delta$ and from Eq.~(\ref{eq:coherence_bound}) we can extract the bound $|\zeta_{0b}^{\rm coh}| \leq C\theta^2N_B\ e^{-\Delta^2/4a^2} /\Delta T.$
Hence for $h\sim \mathcal{O}(\Delta)$, coherences are roughly a factor of $\theta^2 N_B/\Delta T$ smaller than the low-lying populations. In the special case where the ground state manifold becomes degenerate, we instead have the weaker bound $|\zeta^{\rm coh}_{00'}| \leq C\theta^2 |\Delta G_{00'}|\beta /T$ --- thus if the matrix element $\Delta G_{00'}$ is finite, we may have large coherence between the groundstates whenever $\theta^2 \beta/T$ is not small. 

\textbf{High temperature:} At high temperatures, $\Delta\beta  \ll 1$, we can use the coherence bound (setting $\epsilon_b > \epsilon_a$) $|\zeta^{\rm coh}_{a\neq b}| \leq C' \theta^2\beta  N_B  e^{-\beta \epsilon_a}/TZ$, where $Z$ is the partition function. We see that coherences are suppressed by a small factor of $\theta^2\beta N_B/T$ relative to the upper bound one would obtain from simply applying the Cauchy-Schwarz inequality, $|\sigma_{ab}|\leq \sqrt{p(\epsilon_a)p(\epsilon_b)}\leq  p(\epsilon_a)$. The vanishing of coherences in the high temperature regime is consistent with previous works studying thermalisation in the case of physical baths \cite{lee2022perturbative, timofeev2022hamiltonian}. For the populations, since mixing times are typically short at high temperatures \cite{rouze2025efficient, rouze2024optimal, bakshi2024high}, we also expect the corrections to be perturbatively small. 

We finish this section with two comments: firstly, we note that while the mixing time enters the worse-case bound in Eq.~(\ref{eq:tracebound}), from the splitting of Eq.~(\ref{eq:Dyson}) it is clear that the physically relevant timescale is the time for the off-diagonal perturbation $\hat \xi$ to relax. While it is upper bounded by the mixing time, it may be much shorter if $\hat \xi$ does not couple directly to the slowest mixing modes; we provide further discussion in Section \ref{sec:discussion}. If coherences are suppressed and this timescale remains short, we expect to be able to prepare the Gibbs state accurately in practice. 

Second, we point out that the steady state corrections described above can be understood from a complementary point of view, by rewriting $\hat \sigma$ in the form of a modified Gibbs state
$\hat \sigma \propto e^{-\beta \hat H_S'}$, defined with respect to a `renormalised' system Hamiltonian $\hat H'_S = \hat H_S+\theta^2\hat C$. We provide a derivation of the $\hat C$ term in Appendix~\ref{app:renormalisation}, where we show how $\hat\sigma$ can be understood as the fixed point of a detailed balanced Lindbladian in the interaction frame defined with respect to $\hat H'_S$. The term $\hat C$ leaves the diagonal part unspecified, reflecting the ambiguity in fixing the populations at leading order in $\theta$: we note that establishing desirable properties of the perturbation $\hat C$ such as quasilocality appears intimately connected to bounding $\hat \zeta^{\rm pop}$ efficiently.

\section{Numerical results}\label{sec:numerics}

In this section, we present numerical results for the cooling protocol, in order to confirm the arguments presented in Section \ref{sec:steadystates}. As a warm-up, we first present results for single spin cooling, illustrating the perturbative formulas derived in Section \ref{sec:thermal}. Second, we study the practically relevant case of the 2D Quantum Ising model, where we show accurate cooling for a range of system parameters. In Appendix \ref{app:fermions}, we provide a systematic study of cooling for a 1D free fermion chain, where we are able to test the protocol for large system sizes and confirm the perturbative arguments of Section \ref{sec:thermal}.

\subsection{Single spin cooling}\label{sec:singlespin}

 \begin{figure}[h!]
    \centering
    \includegraphics[width=.98\linewidth]{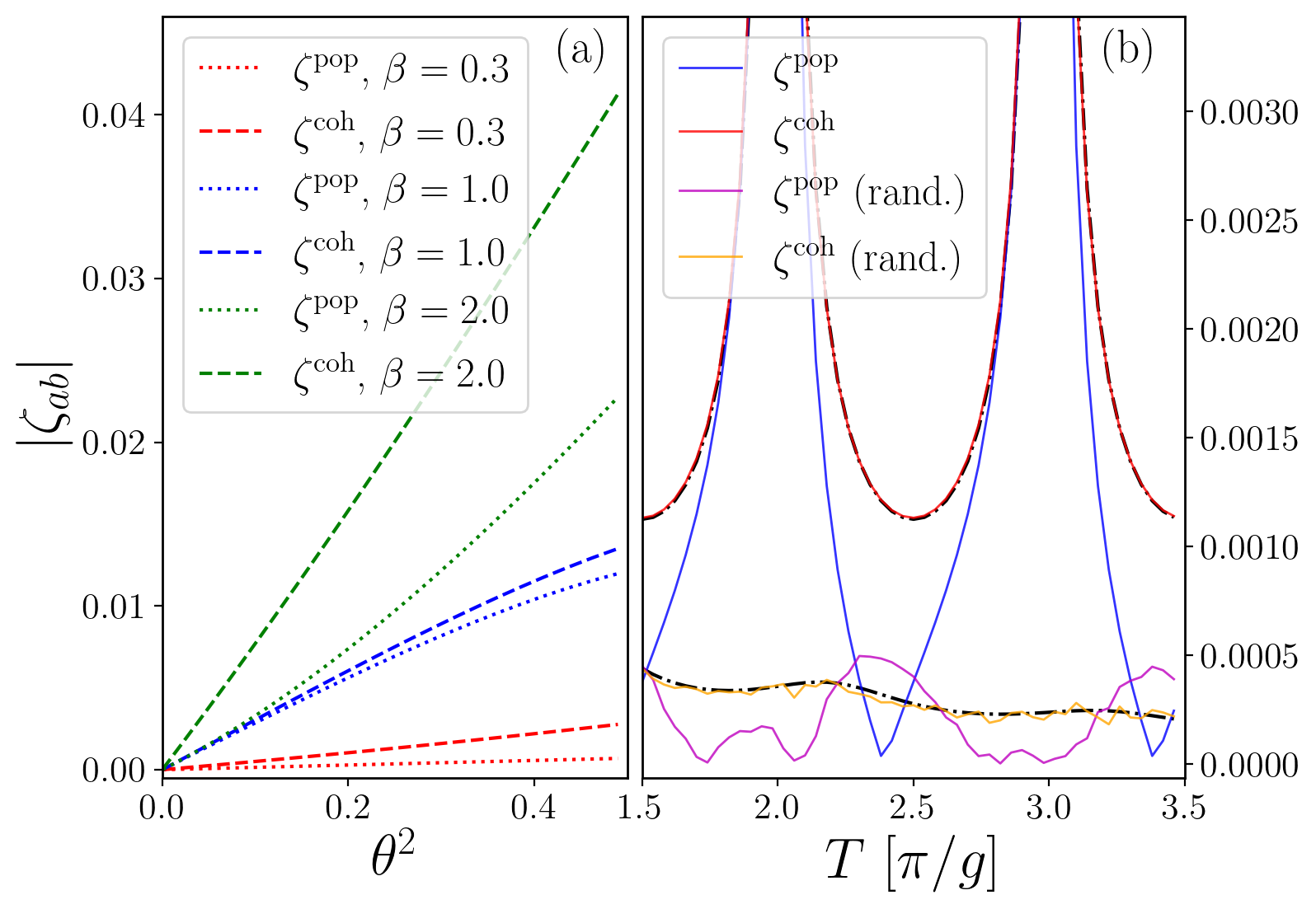}
    \caption{\emph{Single spin cooling}. (a) Scaling of steady state population errors ($\zeta^{\rm pop}$) and coherences ($\zeta^{\rm coh}$) vs.~system-bath coupling $\theta^2$. (b) Comparison of randomized (solid orange, purple lines) and unrandomized (solid red, blue lines) protocols, for varying reset time $T$ and $\theta=0.25$. Black dot-dashed lines show perturbative solution for coherences, Eq.~(\ref{eq:corrections}).}
    \label{fig:spin}
\end{figure}

To support the validity of the perturbative analysis in Section \ref{sec:thermal}, and compare the performance of randomized and unrandomized protocols, we simulate the cooling of a single spin with the Zeeman Hamiltonian 
\begin{equation} 
\hat H_S = - \frac{g}{2}\hat Z.
\end{equation} 
We set $g=1$ throughout. The spin is coupled to a single auxiliary with parameter $h=g$, and simulations performed by exact real-time evolution. 

The steady state corrections are shown in Figure \ref{fig:spin}. In Fig.~\ref{fig:spin}a, we plot the matrix elements of the correction $\hat \zeta = \hat\sigma-\hat\sigma_{\beta}$, as a function of the coupling $\theta$. We denote the magnitude of the matrix elements by $|\zeta_{00}| = \zeta^{\rm pop}$ and $|\zeta_{01}| = \zeta^{\rm coh}$. The results shown are for the unrandomized protocol with three different protocol temperatures. The predicted scaling $|\zeta_{ab}| \propto \theta^2$ is evident for small $\theta$. 

In Fig.~\ref{fig:spin}b, we compare randomized and unrandomized protocols, as the reset time $T$ is varied. We take $\beta = 1$, $\theta=0.25$ and for the randomized protocol use parameter $\lambda = 1$. The solid coloured lines are the numerical results, obtained by time-averaging the state density matrix between $1000/\theta^2\leq n\leq 3000/\theta^2$ reset cycles. The black dot-dash lines give the perturbative solution for the coherences, Eq.~(\ref{eq:corrections}). The results for the 
unrandomized protocol confirm that the correction diverges at resonant values of the drive period, as predicted from Eq.~(\ref{eq:corrections}) by setting $\lambda = 0$. Interestingly, the perturbative solution for the coherences matches the true solution well even in this regime. Randomizing the protocol effectively suppresses the resonances, for both populations and coherences. While our perturbative solution does not give a prediction for the exact population correction, we observe that the order of magnitude follows that of the coherences, consistent with Eq.~(\ref{eq:population_error}).

\subsection{2D Quantum Ising model}\label{sec:ising}

\begin{figure}
    \centering
    \includegraphics[width=.9\linewidth]{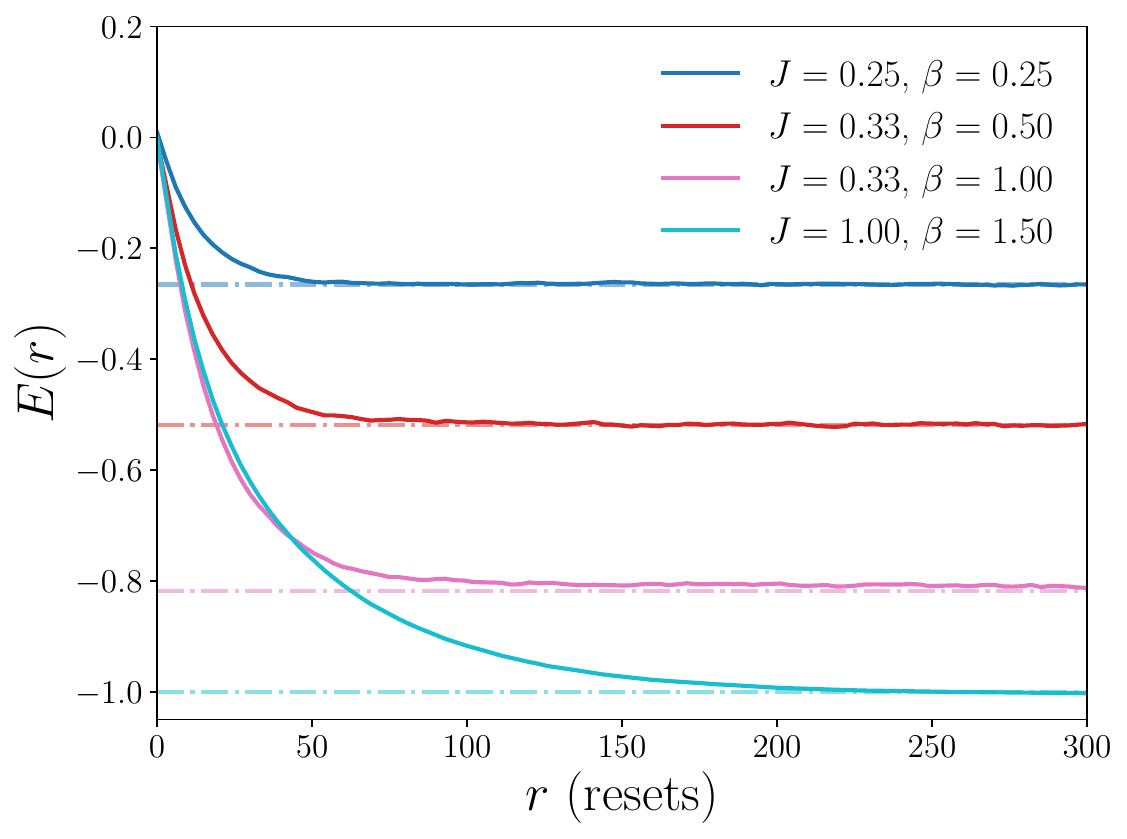}
    \caption{Evolution of energy in 2D Quantum Ising model ($3\times3 $ sites), vs.~number of resets. We show results for several different choices of parameters $J, \beta$, using the unrandomized protocol with $\theta=0.25$. Energies for different curves are scaled to lie between -1 (ground state) and 1 (highest excited state) for comparison. Dot-dashed lines are the corresponding thermal state values.}
    \label{fig:trajectories}
\end{figure}

\begin{figure*}
    \centering
     \includegraphics[width=.96\linewidth]{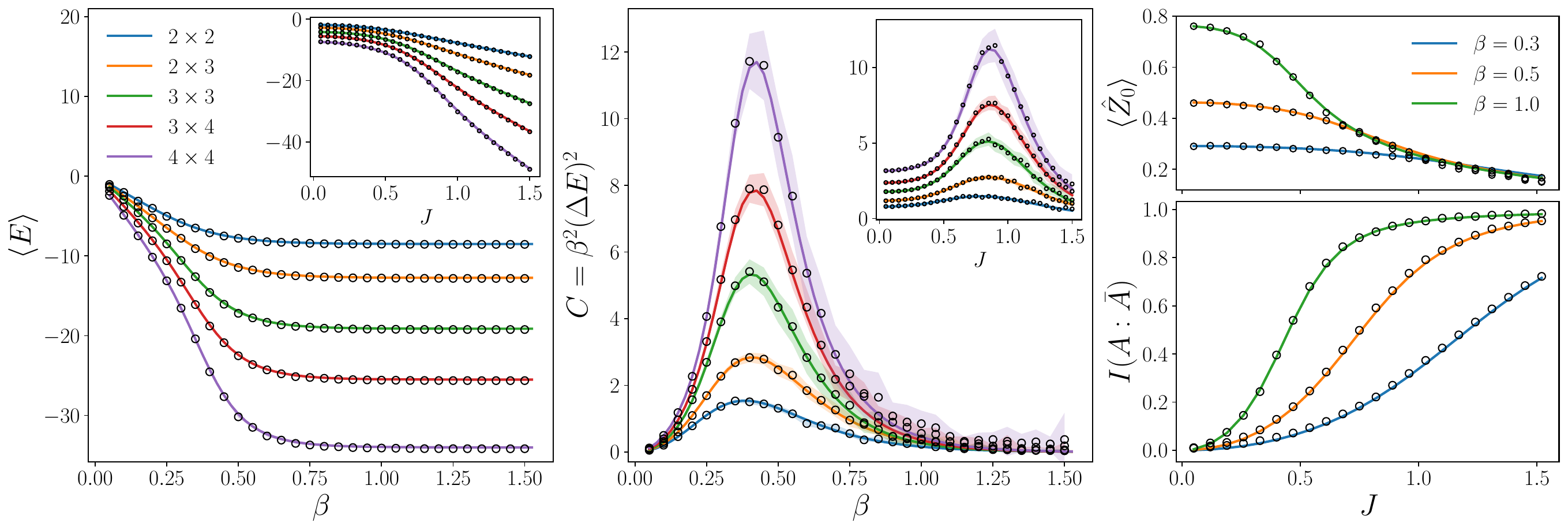}
     \caption{\emph{Steady state observables in 2D Quantum Ising model.} Numerical values are depicted by open points, while coloured lines are values for the thermal density matrix obtained via exact diagonalisation. (a) Total energy in steady state for fixed $J=1$ and varying $\beta$, with system sizes between $2\times 2$ and $4\times 4$. \emph{Inset}: Energy for fixed $\beta=0.5$ and varying $J$. (b) Total heat capacity in steady state, for same parameters as in (a) and varying $\beta$. Error bars are due to finite sampling over trajectories. \emph{Inset}: Heat capacity for fixed $\beta=0.5$ and varying $J$.  (c) {Upper plot}: Transverse magnetisation $\langle \hat Z_0\rangle$ for three different temperatures and varying $J$. {Lower plot:}  Mutual information between a single spin and the rest of the system. System size is $3\times 3$ sites.}\label{fig:observables}
\end{figure*}

We now turn to the question of thermal state preparation in a many-body model. As our example, we take the interacting two-dimensional square-lattice quantum Ising model, and consider lattice sizes of up to 16 spins. The Hamiltonian is defined for a lattice of spin-$1/2$ degrees of freedom as 
\begin{equation}\label{eq:Hising}
    \hat H_{\text{QI}} = -J\sum_{\langle i,j\rangle} \hat X_i\hat X_j -g\sum_i\hat Z_i,
\end{equation}
where the couplings run over all nearest-neighbour pairs, and for $J>0$ the spins favour ferromagnetic alignment.
We will focus on the digital cooling protocol with the corresponding Floquet unitary,
\begin{equation}
    \hat U_\text{QI} = \exp\Big(i\delta g\sum_i \hat Z_i\Big) \exp\Big(i\delta J \sum_{\langle i,j\rangle} \hat X_i \hat X_j\Big),
\end{equation}
and periodic boundary conditions assumed. The model is non-integrable and hosts (in the thermodynamic and prethermal limits) two quantum phases, a quantum paramagnetic phase at small $J$ and high temperature, and a ferromagmetic ordered phase at large $J$ and low temperature. The phase transition is known from Monte Carlo computations \cite{hesselmann2016thermal}, with a thermal phase transition for $g = 0$ at the critical temperature $\beta_c \approx 0.44 J^{-1}$, and a quantum phase transition for $\beta\to\infty$ at $J_c \approx 0.33 g$. In the rest of this section we set $g=1$ and study cooling as parameters $J$ and $\beta$ are varied.

We use the quantum trajectory method to simulate the cooling dynamics in real time, using Google's qsim quantum simulator package \cite{isakov2021simulations}. Initial states are sampled uniformly from the set of computational basis states. We fix a small Trotter angle $\delta=\pi/40$ in order to minimise errors due to the Floquet driving. We found that taking a larger angle $\pi/20$ did not significantly change our results. As discussed in Section \ref{sec:protocol}, the bath parameter $h$ should be chosen to approximately target low-energy transitions of the system, and we fix $h = \text{max}(2g,4J)$ which ensures transitions are in resonance with low-lying excitations. Due to the computational overhead involved, we use only a small number of auxiliaries ($N_B \leq 4$); to maintain translational symmetry we choose the geometry of the system-auxiliary couplings $(\mu,i_\mu)$ randomly at the start of each cooling cycle, with a coupling operator $\hat A_\mu = (\hat Z_{i_\mu}+\hat Y_{i_\mu})/\sqrt{2}$. This randomization differs between independently sampled trajectories. We do not expect this randomization to give different results compared to the case where the system is coupled to a finite density of auxiliaries in the bulk (assuming of course energy is transported efficiently).

As a first test, we examine the convergence of the system energy toward the thermal expectation as the number of resets is increased. Results are displayed in Fig.~\ref{fig:trajectories}, for a $3\times3$ lattice and several representative points in the phase diagram. Each curve is obtained by averaging measurements over $1000$ independent trajectories, with a coupling parameter $\theta=0.25$ and $N_B=3$ auxiliaries. We observe that the measured energy accurately converges toward the thermal value at late times. We do not observe any significant difference between the equilibration times of different phase points -- however, it is likely that the system size is too small to resolve this question, as coarsening dynamics is naively expected for large systems cooled into the ferromagnetic phase \cite{bray1994theory}.  

Next, we turn to an analysis of the steady state. The steady state distribution is accessed by sampling over late time configurations, analogous to classical Monte Carlo schemes. We define the steady state density matrix as 
\begin{equation}\label{eq:steady_state_ensemble}
    \hat \sigma = \mathbb{E}[\ket{\varphi_k}\bra{\varphi_k}] = \frac{1}{K} \sum_{k=1}^K \ket{\varphi_k}\bra{\varphi_k},
\end{equation}
where $\varphi_k$ represents the trajectory wavefunction sampled periodically in time.

In Figure~\ref{fig:observables}, we show the expectation values for different steady state observables, as a function of both $\beta$ and $J$. In each case the open points represent the data from the steady state simulations, while solid coloured lines are the exact thermal values. We show data for lattice sizes of up to $4\times4$ system sites. In each case, we fix the coupling $\theta$ according to $\theta^2 = 0.05 /\sqrt{\beta h}$ \footnote{This scaling choice is motivated by taking the limit of $\omega \to0$ in the formula for the coherences corrections (\ref{eq:corrections})}.

In Fig.~\ref{fig:observables}a, we compare the total system energy. In the main panel we fix $J=1$ and allow $\beta$ to vary, while the inset shows results for fixed $\beta = 0.5$ and varying $J$. In both cases we observe excellent agreement with the exact results (error bars are smaller than the line width). 

Further, in Fig.~\ref{fig:observables}b, we illustrate the system heat capacity, defined as $C=\beta^2(\Delta E)^2$, where $(\Delta E)^2$ is the variance in the energy. We show results for the same parameter values as in Fig.~\ref{fig:observables}a. The data closely follows the exact curve, with observable error bars due to finite sampling of trajectories. A characteristic peak in the heat capacity clearly develops for larger system sizes, close to the critical values of $\beta$ and $J$. This suggests that the thermal cooling protocol is capable of preparing critical many-body states. 

The total energy and the heat capacity are both diagonal in the Hamiltonian eigenbasis and only probe the population errors. Furthermore, the close agreement of these local observables does not guarantee that correlation functions between spatially separated degrees of freedom are accurately represented. Next we consider two measures which are also sensitive to off-diagonal corrections: the on-site transverse magnetisation $\langle \hat Z_0\rangle$, and the mutual information between a single spin (denoted $A$) and the rest of the system ($\bar{A}$): $I({A:\bar{A}}) = S(\hat \rho_A)+S(\hat \rho_{\bar{A}})-S(\hat \rho)$, where $S(\hat\rho) = -\text{tr}(\hat\rho\log \hat \rho)$ is the von-Neumann entropy.
The mutual information is a good measure of the `total' correlation between two systems, since it upper bounds the connected correlation functions of arbitrary observables supported on $A$ and $B$ \cite{wolf2008area},
\begin{equation}
    I(A:B) \geq \frac{\langle \hat O_A \hat O_B\rangle^2 _c}{2 \|\hat O_A\|^2\|\hat O_B\|^2}.
\end{equation}
For pure quantum states, the bipartite mutual information $I({A:\bar{A}})$ reduces to (twice) the entanglement entropy, while for mixed states it also captures classical correlations. Note that the mutual information is a function of the full density matrix and we expect it to capture the error relevant for local correlation functions. In Figure \ref{fig:observables}c, we compare $\langle \hat Z_0\rangle$ and $I({A:\bar{A}})$ for the steady state density matrix and the thermal value from exact diagonalisation. We show the curves for a range of $J$ and three different values of $\beta$, using a system size of $3\times3$ qubits. The steady state shows good agreement with the thermal state values in all cases.

To conclude our discussion of the 2D Ising model, we check how the errors to the Gibbs state scale with the coupling parameter $\theta$. We focus on two measures of the protocol accuracy: the relative error in the steady state energy, and the trace distance between the steady state and the Gibbs state. The trace distance between two state $\hat\sigma$ and $\hat\rho$ is  defined as
\begin{equation}
    D(\hat\sigma,\hat\rho)=\frac{1}{2}\|\hat\sigma-\hat\rho\|_1 \equiv \frac{1}{2}\text{tr}\sqrt{(\hat\sigma-\hat\rho)^2},
\end{equation}
and takes values in the range $[0,1]$. When computing the trace distance between the steady state and the Gibbs state, we observed that the dominant error at small $\theta$ comes from the finite Trotter angle in our numerics (precisely, we observe an offset of $D(\hat\sigma,\hat\sigma_\beta)\approx 0.1$ in the limit of small $\theta$, for $\delta = \pi/40$). In order to isolate the dependence on the system bath coupling, we compute the trace distance $D(\hat\sigma,\hat \sigma_\beta^{(1)})$ with respect to the Gibbs state of the first order Floquet Hamiltonian \cite{yin2023prethermalization}, $\hat \sigma_\beta^{(1)} \propto e^{-\beta H_F^{(1)}}$, where
\begin{equation}
    \hat H_{F}^{(1)} = \hat H_{\rm QI} + \frac{i\delta}{2} \big[\hat H_J, \hat H_g],
\end{equation}
and $\hat H_J$ and $H_g$ are the first and second terms in Eq.~(\ref{eq:Hising}) respectively. Similar results can be obtained by decreasing the Trotter angle in the simulations and computing $D(\hat\sigma,\hat\sigma_\beta)$, at the cost of increased simulation time. The result does not improve significantly for higher-order Floquet Hamiltonians.

\begin{figure}
    \centering
    \includegraphics[width=0.9\linewidth]{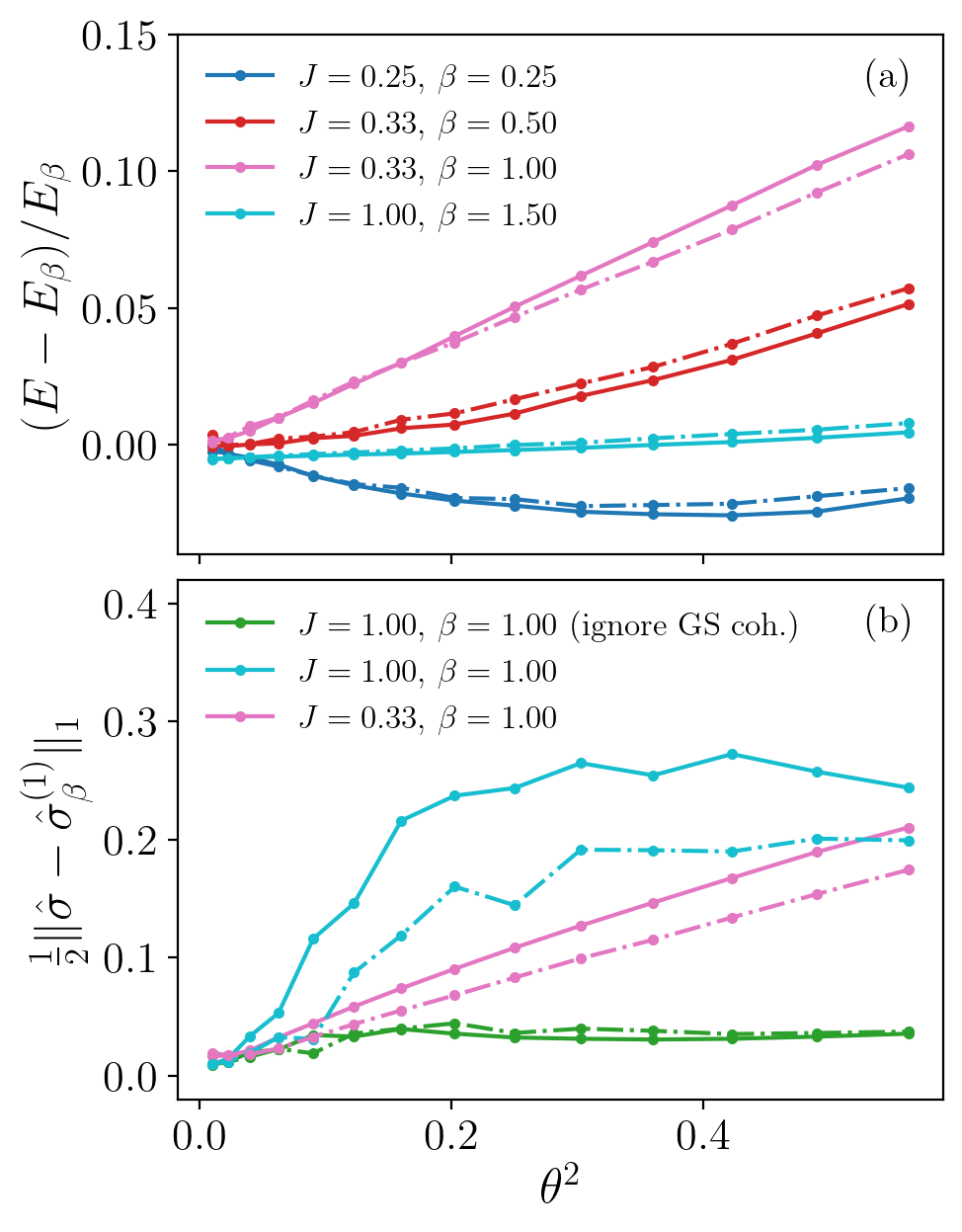}
    \caption{(a) Scaling of relative error in steady state energy, relative to thermal energy, vs.~coupling $\theta^2$. We show four representative points in the $J/\beta$ phase diagram. Solid lines are for the unrandomized protocol, while dot-dash lines are for the randomized protocol with parameter $\lambda = 5$. (b) Scaling of trace distance between steady state and the first order Floquet Gibbs state, vs.~coupling $\theta^2$. We focus on the two low-temperature points, and for $\beta=1.5$, $J=1$ show also the value when the coherence between the degenerate ferromagnetic ground states is ignored.}
    \label{fig:errors}
\end{figure}

We show results as a function of the coupling $\theta^2$, for different values of $J$ and $\beta$ in Fig.~(\ref{fig:errors}); each data point is averaged over 40000-100000 measurements, sampled from the cooling trajectory every 10 resets. We use a system size of $3\times 3$, with 3 auxiliaries. The errors in both the energy and the trace distance are expected to depend on the system size, however our small system numerics are not sufficient to determine this scaling. The scaling will vary depending on the model in question. We also compare the unrandomized (solid lines) and randomized (dot-dash lines) protocols, with a randomization parameter $\lambda =5$. 

The relative error in the energy (Fig.~\ref{fig:errors}a) converges cleanly toward zero as the strength of the coupling parameter $\theta$ is reduced, with the deviations largest near the critical value of $J=0.33$, $\beta = 1$. For the trace distance (Fig.~\ref{fig:errors}b), we show only the two low temperature points $\beta = 1$, $J=0.33$, and $\beta = 1.5$, $J=1$, as the higher temperature points converge too slowly in the number of samples (the Monte Carlo method is not ideally suited to computing this quantity).  We observe in both cases a convergence toward zero as the coupling strength is reduced. Deep in the ferromagnetic phase, on account of the Ising $\mathbb{Z}_2$ symmetry, the Hamiltonian ground state is degenerate with an exponentially small splitting in the system size. In this case, we expect potentially large coherence between the two ground states. By subtracting the contribution from ground state coherence to the trace distance at the ferromagnetic point $\beta = 1.5$, $J=1$, we observe a significantly closer agreement to the Gibbs state. Note that in the thermodynamic limit, due to e.g.~external noise sources, we expect the Ising symmetry to be spontaneously broken and the system will be found in one of the two symmetry-breaking states.  

For the energy error, we observe that the errors in randomized and unrandomized protocols are essentially the same at small values of $\theta$. For the trace distance, the randomized protocol leads to a more significant decrease in the errors, but both protocols perform well in the limit of small $\theta$. It is possible that for the system sizes we access, the resonances appearing in the perturbation theory of Section \ref{sec:thermal} do not play a strong role. For the free-fermion model studied in Appendix \ref{app:fermions}, where we can access significantly larger system sizes, the effect of resonances appears clearly, although the errors in both randomized and unrandomized protocols remain small in the weak-coupling limit.

Our numerical results confirm the perturbative picture developed in Section \ref{sec:steadystates} and show that the modulated coupling protocol accurately prepares thermal states in many-body, interacting quantum systems. Furthermore, we point out that the protocol parameters used in this section are modest in that the cooling cycle involves a relatively small number of unitary layers. For example, for $J=g=1$, $\beta \approx 0.5$, which corresponds to the preparation of the `critical state' in the main panel of Fig.~\ref{fig:observables}b, we used a coupling parameter of $\theta \approx 0.2$, with a reset time $T = 3a^{-1}\approx 0.7$ (or $M = 9$ for $\delta = \pi/40$), and $N_B=3$ auxiliaries for the $4\times4$ system. These values are within easy reach for future experiments on current hardware. Since the reset time needs only scale as $T \propto \beta^{1/2}$, preparation of low-temperature and ground states also does not pose a fundamental issue. 

\section{Discussion } \label{sec:discussion}

As we have seen in Sections \ref{sec:protocol} and \ref{sec:numerics}, the accuracy of our protocol improves as the magnitude of the system-bath coupling $\theta$ is lowered, relative to the energy scales of the system Hamiltonian. This is due inherently to the fact that our protocol is defined in the Schrödinger picture, as discussed in Section \ref{sec:thermal}. In contrast, other proposals for exact ~\cite{chen2023efficient, gilyen2024quantum} or approximate \cite{chen2023quantum, hahn2025provably, ding2025end} quantum Gibbs samplers have worked primarily at the level of the Lindbladian in the interaction picture, i.e.~Eq.~(\ref{eq:lindblad}). Similar to our work, the protocol of Ref.~\cite{ding2025end} works in the Schrödinger picture, but their bounds do not appear to utilise this fact beyond determining the validity of the Lindblad equation. 

The accuracy of these protocols is instead controlled by 
applying the mixing time bound Eq.~(\ref{eq:mixing}), which gives the error to the Gibbs state $\|\hat \zeta\|_1 \lesssim t_{\rm mix}\|[\Delta \hat G,\hat \sigma_\beta ]\|_1 $. As we discussed in Section \ref{sec:thermal}, $t_{\rm mix}$ is an intrinsic timescale of the thermalisation dynamics which is approximately independent of the coupling $\theta$. Rather, the error is made small by increasing the Gaussian filter width $a^{-1}$, which in turn allows to bound the size of the commutator term $\|[\Delta \hat G,\hat \sigma_\beta ]\|_1$ more strongly. For example, in Ref.~\cite{hahn2025provably}, the authors rigorously showed the errors are upper bounded by $\|\hat{\sigma}-\hat \sigma_\beta\|_1 \leq \tilde{\mathcal{O}}(a \beta N_B t_{\text{mix}})$, where $\tilde{\mathcal{O}}$ denotes an upper bound neglecting poly-logarithmic factors. In this work, we argued (without establishing a rigorous bound on the trace-distance error) that even when the commutator norm is $\mathcal{O}(N_B)$, we can systematically control the error by reducing $\theta$. 

It is easy to see that our perturbation theory fails if one tries to solve for the steady state in the interaction picture: since there is no $\mathcal{O}(1)$ term in the Lindblad equation Eq.~(\ref{eq:lindblad}), there is no sense in which $\theta^2$ can be treated as a small-coupling parameter. We can approach the interaction picture map smoothly by applying an additional `rewinding' step at the end of each reset cycle, as is implemented in the algorithm of Ref.~\cite{hahn2025provably}: after applying each reset cycle, Eq.~(\ref{eq:resetchannel}), we apply a reverse time evolution of duration $2T^{*}$ i.e.~apply the unitary $e^{+2iT^{*}\hat H_S}$. The interaction picture map corresponds to a full rewinding $T^{*} = T$. Accounting for the rewinding in our perturbation theory, we neglect the randomization step ($\lambda = 0$) and obtain for the coherences: 
\begin{equation}
    \zeta^{\rm coh}_{a\neq b} = i\theta^2(\Delta G)_{ab}\bigg(\frac{p(\epsilon_b) - p(\epsilon_a)} {1 - e^{2i(T-T^{*})\omega_{ab}}}\bigg). \label{eq:rewind}
\end{equation}
The errors diverge as $\hat \zeta_{a\neq b}^{\rm coh} \propto (T-T^{*})^{-1}$ as we approach the interaction picture channel. Therefore, while it appears harder to rigorously bound the accuracy of protocols in the Schrödinger picture, protocols defined in the interaction picture miss a physically important effect coming from the dephasing of coherences due to Hamiltonian evolution in the Schrödinger picture. It would be desirable to extend and potentially strengthen rigorous bounds as used in \cite{chen2023quantum,  hahn2025provably, ding2025end} to protocols like the one considered here. In protocols where the smallness of the error is controlled by a different parameter than $\theta$, as in e.g.~\cite{hahn2025provably, ding2025efficient}, we expect the combining these approaches with our weak-coupling Schrödinger map approach can lead to more accurate state preparation.

Finally we remark that in all of the proposed Gibbs sampling protocols, including ours through our discussion of the population errors, the mixing time $t_{\rm mix}$ enters since it controls the worst-case decay time of arbitrary errors. This presents a pessimistic view of state preparation: many phases of interest are expected to feature mixing times growing exponentially with the system size, due to the phenomena of spontaneous symmetry breaking. Nevertheless, our discussion in Section \ref{sec:thermal} makes it clear that the relevant timescale should be set by the decay of the specific error $\hat \xi \propto \mathcal{L}(\hat\sigma_\beta)$. As we argued, this error decays quickly in part due to dephasing of coherences. In the 2D Ising model, the magnetisation order parameter $\hat M = \sum_i \hat X_i$ is an example of an operator which takes an exponentially long time to relax to its equilibrium value (zero by symmetry). Generic operators which develop a finite overlap with the magnetisation during their dynamics are therefore also expected to decay exponentially slowly. However, if the dynamics is designed to preserve the Ising $\mathbb{Z}_2$ symmetry, then  the operator $\mathcal{L}(\hat\sigma_\beta)$ is blocked from developing any overlap with the magnetisation: $\hat M$ is odd under the $\mathbb{Z}_2$ symmetry, while $\hat \sigma_\beta$ and $\mathcal{L}(\hat\sigma_\beta)$ are even. Indeed many operators e.g.~total energy exhibit coarsening dynamics at low temperatures, decaying to equilibrium on timescales which grow only polynomially with system size \cite{bray1994theory}. A careful consideration of protocol symmetries and operator decay is likely to be important in efficiently preparing non-trivial phases. 

\section{Conclusions}\label{sec:conclusion}

In summary, we have presented an algorithm for preparing Gibbs states of quantum many-body Hamiltonians. Our protocol relies on engineering Lindblad evolution using auxiliary qubits, which {\it approximately} satisfies the quantum detailed balance condition, thereby driving the system towards the Gibbs state. Compared to recently proposed exact Gibbs samplers~\cite{chen2025efficient, ding2025efficient, gilyen2024quantum} which require significant resources for implementation on quantum hardware, our algorithm offers a straightforward and efficient path to near-term experiments. Indeed, a closely related algorithm for ground state preparation in correlated systems has been recently demonstrated with a superconducting quantum processor~\cite{mi2024stable}. 

While this paper focused on establishing a theoretical basis for approximate Gibbs state sampling, a promising future direction is to investigate cooling dynamics and mixing times in specific examples of correlated phases, especially in two dimensions, as well as to evaluate the potential of dissipative-engineering algorithms for quantum chemistry applications in molecular systems. Our results for the 2D Ising model, though limited to modest system sizes, show the potential for thermal state preparation of non-trivial quantum phases and critical points. In this setting, many open questions exist concerning the timescales involved in state preparation and the stability to breaking of detailed balance. So far, rigorous results on mixing times have been limited to high temperature systems \cite{rouze2025efficient, rouze2024optimal, bakshi2024high}, or weakly-interacting spin and fermion systems \cite{tong2024fast, vsmid2025polynomial, zhan2025rapid}. Despite the usefulness of these results, it is unlikely that quantum advantage can be found in these settings. Understanding the fate of Gibbs samplers in preparing highly correlated quantum states will require further efforts and insight on a case-by-case basis, as in classical Monte Carlo applications \cite{levin2017markov}. 

A further open question concerns the stability of cooling protocols to the presence of unwanted device noise. In experimental cases, noise which couples directly to populations (which relax to equilibrium only on the perturbative timescale $t_\theta \sim 1/\theta^2$) is likely to be the major barrier to accurate state preparation. We expect the errors in the presence of noise to scale approximately as $\sim \gamma/\theta^2$, with $\gamma$ a measure of the integrated noise over one reset cycle. 
For noise rates $\gamma$ exceeding the bound $\gamma /\theta^2 \gg 1$, the protocol will fail \cite{qpcooling}. However, this naive estimate does not take into account the stability of many ordered phases (e.g.~2D Ising model) to certain types of noise. A more complete understanding of how weak noise affects the detailed balance relations, and the ensuing stability of quantum state preparation algorithms, is required. 

Turning to experiments, one immediate application will be to simulate phases and quantum-critical points of spin models in 2d, such as the quantum Ising model considered above and previously studied experimentally on a digital processor in Ref.~\cite{mi2024stable}, as well as XY and Heisenberg models in the analog-digital setting~\cite{AndersenNature2025Analog}. The latter case may allow for reaching lower temperatures due to reduced effects from environmental noise. Furthermore, exploring realizations of cooling algorithms for many-body Gibbs state preparation in other platforms such as neutral atom arrays \cite{henriet2020quantum, bluvstein2024logical} and trapped ions \cite{monroe2021programmable, foss2024progress}, would be desirable. While early proposals exist for simulating open quantum systems on analog platforms \cite{diehl2008quantum, barreiro2011open}, it would be timely to revisit these questions in light of the recent progress in efficient thermal state preparation. 

\acknowledgments

We thank Dominik Hahn and Chi-Fang Chen for useful discussions, and Alexios Michailidis, Xiao Mi, and Vadim Smelyanskiy for a collaboration on related work. 
This work was partially supported by the European Research Council via Grant Agreement TANQ 864597. 

\appendix

\begin{widetext}

\section{Thermal state preparation for free-fermion chains }\label{app:fermions}

\begin{figure}
    \centering
     \includegraphics[width=.96\linewidth]{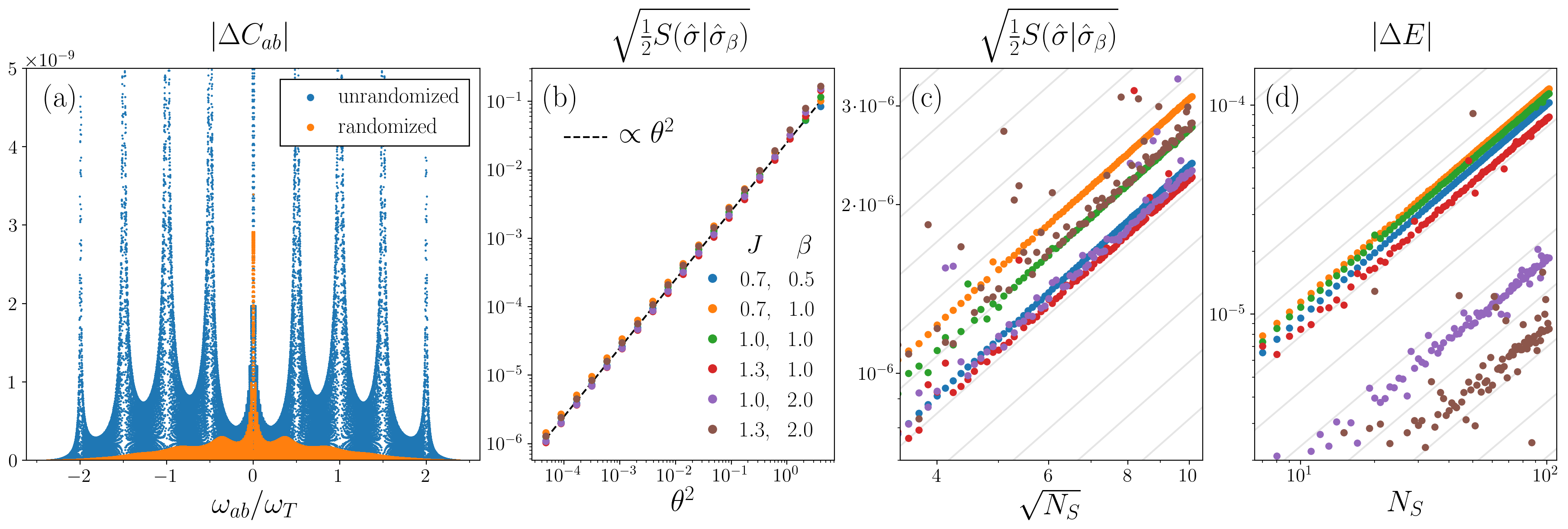}
     \caption{Thermal state preparation for the free fermion chain, Eq.~(\ref{eq:fermionHamiltonian}) \textbf{(a)} Absolute value of the steady state covariance matrix elements in the energy basis, with the thermal state populations subtracted. We plot the matrix elements as a function of the frequency $\omega_{ab}/\omega_T$, where $\omega_{ab} = \epsilon_a-\epsilon_b$ are the single-particle Bohr frequencies and $\omega_T = 2\pi/T$ is the drive frequency. \textbf{(b)} Square root relative entropy between steady state and Gibbs state, as a function of coupling strength $\theta$. \textbf{(c)} Square root relative entropy between steady state and Gibbs state, as a function of system size. \textbf{(d)} Error in the total steady state energy, as a function of system size. In \textbf{(c)} and \textbf{(d)}, grey lines denote the scaling $y(x) \propto x$. }
\label{fig:fermions}
\end{figure}

In this Appendix, we study the cooling protocol for the example of the Majorana free-fermion chain, 
\begin{equation}\label{eq:fermionHamiltonian}
    \hat H_S = \frac{ig}{2}\sum_{j=0}^{N_S-1} \hat\gamma_{2j}\hat\gamma_{2j+1} - \frac{iJ}{2}\sum_{j=0}^{N_S-2} \hat\gamma_{2j+1}\hat\gamma_{2j+2},
\end{equation}
where the Majorana fermions satisfy $\{ \gamma_i,\gamma_j\}=2\delta_{ij}$.
Equivalently, this is the quantum Ising chain Hamiltonian after Jordan-Wigner transforming. Defining the vector of Majoranas, $(\vec{\gamma})_k \equiv \hat \gamma_k$, the Hamiltonian can be written as $\hat H_S = \frac{i}{4}\vec{\gamma}^TA_S\vec{\gamma}$, where $A_S$ is a $2N_S\times 2N_S$ matrix with elements defined by Eq.~\ref{eq:fermionHamiltonian}. The matrix $A_S$ is real and antisymmetric: it can be brought into canonical form by an orthogonal transform $O_A$, with $\vec{\alpha} = O_A\vec{\gamma}$
\begin{equation}
    \hat H_S = \frac{i}{4}\vec{\alpha}^T h_S\vec{\alpha}, \hspace{1cm} h_S = \bigoplus_{k=0}^{N_S-1} \begin{pmatrix}
        0 & \epsilon_k \\ -\epsilon_k & 0
    \end{pmatrix},
\end{equation}
where $\epsilon_k$ are the single-particle energies (eigenvalues of $iA_S$). For the Hamiltonian Eq.~(\ref{eq:fermionHamiltonian}), the energies are given by \cite{lieb1961two}
\begin{equation}
    \epsilon_k = \sqrt{J^2+g^2+2gJ\cos(\phi_k)},
\end{equation}
where in the limit of a large chain with open boundaries the $\phi_k \approx \pi k/N_S$. The spectrum is gapped, except at the critical point $J=g$ where the gap closes at $\phi_k = \pi$. 

Due to the Gaussian nature of the free fermion state, the cooling protocol can be implemented exactly numerically for large system sizes  \cite{barthel2022solving, surace2022fermionic, qpcooling}: rather than working with the full quantum wavefunction, the fermionic covariance matrix $(C_\rho)_{ij} = \frac{i}{2}\langle [\hat \gamma_i,\hat\gamma_j]\rangle_\rho$ is propagated. So long as $\hat\rho$ remains a Gaussian fermionic state, which is true as long as the system Hamiltonian is quadratic in fermions, and the jump operators are linear in fermions, 
then due to Wick's theorem the covariance matrix contains the full information of the many-body state \cite{surace2022fermionic}. This is the main advantage of working with the free fermion model. We also expect our simulations to capture the essential physics of weakly-interacting fermion systems. 

We implement the cooling protocol described in Section~\ref{sec:protocol} for the model Eq.~(\ref{eq:fermionHamiltonian}). We refer to Ref.~\cite{qpcooling} for details of the numerical implementation. For the system evolution, we work directly with the unitary 
\begin{equation}
    \hat U_S = \exp(-i\delta \hat H_S),
\end{equation}
which minimizes the errors due to the Trotter angle $\delta$. We consider cooling operators which are linear in the fermion operators, $\hat A_s = (\hat\gamma_{2s}-\hat \gamma_{2s+1})/\sqrt{2}$. We note that while our choice of cooling operators may be natural in native-fermionic simulation platforms, encoding on spin-qubit platforms requires implementation of a non-local Jordan-Wigner string operator for sites in the bulk of the chain. However the linearity is necessary in order to preserve the Gaussian state property. 

The dephasing channel due to the randomization step, Eq.~(\ref{eq:averagedR}), is implemented directly by scaling the matrix elements of the covariance matrix in the eigenbasis of $A_S$:
\begin{equation}\label{eq:fermion_dephasing}
    C_{ab} \longrightarrow \frac{C_{ab}}{1-i\lambda T(\epsilon_a + \epsilon_b)},
\end{equation}
where $\epsilon_a$ are the eigenvalues defined above. We point out that while the random evolution under $\hat H_S$ preserves the Gaussianity of the fermion state, explicitly averaging over the randomization time does not, since a mixture of Gaussian states is not necessarily Gaussian. Quantities such as the state entropy, which we compute below in the Gaussian approximation, may therefore deviate from the true state value. Still, Eq.~(\ref{eq:fermion_dephasing}) correctly captures the evolution of quadratic observables and we expect it to capture the essential mechanism of the randomization; our numerics below matches our predictions from Section \ref{sec:thermal} regarding the dephasing of coherences.

We now assess the accuracy of thermal state preparation under our cooling protocol. We restrict to the case where each site of the chain is coupled to one auxiliary site, i.e.~cooling operators $\hat A_s$ with $0\leq s\leq N_S-1$. Throughout we use parameters $\delta = 0.005\pi$, $\theta = 0.01$, $g=1$, $h=2\times \text{max}(g,J)$ and the reset time $T = 10 a^{-1}$,  unless specified otherwise. Here $a$ is the filter function width, Eq.~(\ref{eq:adef}). 

First, we consider the effect of randomization in the protocol. In Section~\ref{sec:thermal}, we showed that at the level of perturbation theory, the randomization was necessary to cure uncontrolled divergences in coherences at half-integer multiples of the drive frequency $\omega_T = 2\pi /T$. We numerically solve for the steady state covariance matrix $C_\sigma$, with and without the dephasing map, Eq.~\ref{eq:fermion_dephasing}, and extract the error with respect to the thermal covariance matrix, 
\begin{equation}
    C_{\sigma_\beta} = -\frac{1}{2}\tanh \bigg(\frac{i\beta A_S}{2}\bigg). 
\end{equation}
In Figure~\ref{fig:fermions}a, we plot the absolute value of the matrix elements of $\Delta C = C_\sigma - C_{\sigma_\beta}$ in the eigenbasis of $iA_S$, as a function of of $\omega_{ab}/\omega_T$, where $\omega_{ab} = \epsilon_a-\epsilon_b$. The errors of the unrandomized protocol are shown by the blue points, and for the randomized protocol by the orange points. We use parameters $g = J = 1$, a system size of $200$ sites with one auxiliary per site, a reset time $T = 15 a^{-1}$ to emphasise the effect of randomization, and a randomization parameter $\lambda =2$. In the unrandomized protocol, we observe clear divergences in the magnitude of the error for coherences at half integer multiples of $\omega_T$, as predicted in Section~\ref{sec:thermal}. These divergences are efficiently removed in the randomized protocol, and the only significant error peak occurs in the population sector, $\omega_{ab} \approx 0$. Our finding suggests that the randomization may be crucial for accurate state preparation with large system sizes. 

Next, we consider the errors as a function of the system-bath coupling strength $\theta$ and the system size $N_S$. In order to quantitatively track the total error, we will use Pinsker's inequality to upper bound the trace distance by the relative entropy between the steady state and the thermal state:
\begin{equation}
    \frac{1}{2}\|\hat\sigma - \hat\sigma_\beta\|_1 \leq \sqrt{\frac{1}{2}S(\hat\sigma|\hat\sigma_\beta)}, \hspace{1cm} S(\hat\sigma|\hat\sigma_\beta) = \text{Tr}(\hat\sigma(\log \hat\sigma - \log \hat \sigma_\beta)).
\end{equation}
The relative entropy can be calculated efficiently in our case. We rewrite the relative entropy in terms of free energy differences \cite{donald1987free}:
\begin{gather}
    S(\hat\sigma|\hat\sigma_\beta) = -S(\hat \sigma) - \text{Tr}(\hat\sigma \log \hat \sigma_\beta ) = -S(\hat \sigma) + \log Z + \text{Tr}(\hat \sigma \hat H_S)\beta \nonumber\\ 
    = \beta(E(\hat \sigma) -\beta^{-1} S(\hat \sigma)) - \beta(E(\hat \sigma_\beta) - \beta^{-1}S(\hat\sigma_\beta)),
\end{gather}
where $E(\hat \rho) = \text{Tr}(\hat H_S\hat\rho)$ is the state energy and $S(\hat\rho) = - \text{Tr}(\hat\rho \log \hat \rho) $ is the von Neumann entropy. Both of these quantities can be computed from the covariance matrix $C_\rho$, assuming $\hat\rho$ is a Gaussian fermion state \cite{surace2022fermionic}:
\begin{equation}
    E(\hat \rho) = \frac{1}{4}\text{Tr}(A_S C_\rho), 
\end{equation}
\begin{equation}
    S(\hat \rho) = - \sum_{k=0}^{N_S-1} \bigg[\Big(\frac{1}{2}+\nu_k\Big) \log \Big(\frac{1}{2}+\nu_k\Big) + \Big(\frac{1}{2}-\nu_k\Big) \log \Big(\frac{1}{2}-\nu_k\Big)\bigg],
\end{equation}
where $\nu_k$ are the non-negative eigenvalues of $iC_\rho$. This allows us to efficiently bound the trace norm error for the cooling protocol. 

In Figure \ref{fig:fermions}b, we show the trace norm upper bound $\sqrt{\frac{1}{2}S(\hat\sigma|\hat\sigma_\beta)}$ as a function of the system-bath coupling strength $\theta$. We fix the system size $N_S = 100$, and use the randomized protocol with $\lambda = 2$. We show data for several representative points in the $J-\beta$ phase diagram; for all points, the error displays an almost perfect agreement with the trace-norm error prediction of Section \ref{sec:thermal}, $\|\hat\sigma -\hat \sigma_\beta\|_1 \lesssim  \mathcal{O}(\theta^2)$. 

In Figures \ref{fig:fermions}c-d, we show the error from the relative entropy, as well as the total energy error $\Delta E = E(\hat\sigma)-E(\hat\sigma_\beta)$, as a function of the system size. We show the same phase diagram points as in Fig.~\ref{fig:fermions}b. At low temperature in the $J\gg g$ phase we observe some pronounced fluctuations in the errors, although we cannot rule out the possibility that this is due to our numerical routine, which can struggle to converge in the presence of degeneracies (as occur for $J>g$). Nevertheless, the system size scaling remains clear from the data. For the relative entropy, we observe the scaling $\sqrt{\frac{1}{2}S(\hat\sigma | \hat\sigma_\beta)} \propto \sqrt{N_S}$. This shows that the trace distance error is sub-extensive in the system size, while the relative entropy error grows extensively, as would be expected if the  population of each fermionic mode deviates from its thermal expectation by an $\mathcal{O}(\theta^2)$ amount. For the total energy error, we observe the scaling $|\Delta E|\propto N_S$, which is again natural to expect from small population errors. Note that this scaling is better than one expects based on the trace-distance error, which would give an upper bound $|\Delta E|\leq N_S^{3/2}$. We note that the mixing time of free fermions evolving under Lindblad dynamics similar to that derived in Section \ref{sec:QDB} was analysed in recent work \cite{vsmid2025rapid} and shown to scale as $t_{\rm mix} \sim \log N_S$, i.e. free fermions mix rapidly (assuming a coupling to the bath on system site). Our numerics supports the hypothesis, in line with the perturbative arguments of Section \ref{sec:thermal}, that the errors for free fermion thermal state preparation scale as
\begin{equation}
    \|\hat\sigma-\hat \sigma_\beta\|_1 \leq C \theta^2 \sqrt{N_S},
\end{equation}
possibly up to a $\log N_S$ correction from the mixing time, and with $C$ a function of the remaining protocol parameters. Furthermore, accurate prediction of local observable densities, e.g.~$E/N_S$, appears to be possible with $\mathcal{O}(\theta^2)$ error. 

\section{Weak-coupling derivation of interaction-picture map}\label{app:expansion}

In this appendix, we provide a step-by-step derivation of the Lindblad form of the interaction-picture map, by keeping leading order terms in the weak-coupling expansion. While the derivation mirrors standard ones in the open systems literature \cite{lidar2019lecture}, the derivation for discrete-time processes is uncommon and we reproduce it here for completeness. 

The interaction-picture map was defined in Eq.~(\ref{eq:interactionmapdef}) in the main text. We remind the reader that the bath state is $\hat\phi = \ket{0}\bra{0}^{\otimes N_B}$.  We expand the unitary $\hat{Q}_I$ entering the interaction-picture map to second order: 
\begin{gather}
\hat{Q}_I = \hat{U}_{SB,I}(M)\ldots\hat{U}_{SB,I}(0)\ldots \hat{U}_{SB,I}(-M) \nonumber \\
= 1-i\theta\delta \sum_{\tau=-M}^{M} \hat {V}_I(\tau) -\frac{\theta^2\delta^2}{2}\sum_{\tau=-M}^{M}\hat {V}_I(\tau)^2 - \theta^2\delta^2\sum_{\tau={-M}}^{M} \sum_{\tau'=-M}^{\tau-1} \hat {V}_I(\tau)\hat {V}_I(\tau') \nonumber \\
= 1-i\theta\delta\sum_{\tau=-M}^{M} \hat {V}_I(\tau) - \frac{\theta^2\delta^2}{2} \sum_{\tau,\tau'=-M}^{M} \mathcal{T}(\hat {V}_I(\tau)\hat {V}_I(\tau')).
\end{gather}
In the last line we introduced the discrete-time time-ordering operator $\mathcal{T}$ which is defined (along with the anti time-ordering operator $\mathcal{T}^*$) by:
\begin{equation}
    \mathcal{T}(O(\tau) O(\tau')) = \begin{cases}
        O(\tau) O(\tau') & \tau > \tau' \\
        O(\tau') O(\tau) & \tau \leq \tau'. 
\end{cases}
\hspace{1cm}
    \mathcal{T}^*(O(\tau) O(\tau')) = \begin{cases}
        O(\tau) O(\tau') & \tau \leq \tau' \\
        O(\tau') O(\tau) & \tau > \tau'.
\end{cases}
\end{equation}
Substituting into the definition of the interaction-picture map, and observing that the linear term in $\theta$ vanishes due to $\text{tr}\big(\tilde V_\tau \hat \phi\big) = 0$, we arrive at the second-order map
\begin{equation}\label{eq:interaction_expansion}
    \frac{\mathcal{E}_I(\hat\rho)-\hat\rho}{\theta^2 \delta^2} \approx \sum_{\tau,\tau'=-M}^{M}\bigg[\text{tr}_B\Big(\hat V_I(\tau')[\hat\rho\otimes\hat\phi]\hat V_I(\tau)\Big) - \frac{1}{2}\text{tr}_B\Big(\mathcal{T}(\hat V_I(\tau)\hat V_I(\tau'))[\hat\rho\otimes\hat\phi]\Big) - \frac{1}{2}\text{tr}_B\Big([\hat\rho\otimes\hat\phi]\mathcal{T}^*(\hat V_I(\tau')\hat V_I(\tau'))\Big) \bigg].
\end{equation}
We now use the bath correlation function 
\begin{equation}
    \text{tr}_B( \hat Y_{\mu,I}(\tau)\hat Y_{\mu',I}(\tau') \hat\phi) = \delta_{\mu,\mu'}e^{ih\delta(\tau'-\tau)}
\end{equation}
and the definition of $V(\tau)$ (Eq.~\ref{eq:Vdef}) to write 
\begin{gather}
    \text{tr}_B(\hat V_I({\tau})\hat V_I({\tau'}) \hat\phi) = f(\tau) f({\tau'})e^{ih\delta(\tau'-\tau)} \sum_{\mu=1}^{N_B}\hat A_{\mu,I}(\tau)\hat A_{\mu,I}(\tau').
\end{gather}
This equation takes a more familiar form after introducing the jump operators
\begin{equation}\label{eq:Lderivation}
    \hat L_\mu = \delta\sum_{\tau = -M}^{M} f(\tau) e^{ih\delta\tau} \hat A_{\mu,I}(\tau).
\end{equation}
The first term entering the square brackets in Eq.~(\ref{eq:interaction_expansion}) then becomes $\sum_\mu \hat L_\mu \hat\rho \hat L^\dagger_\mu$. The time-ordered term can be decomposed into Hermitian and anti-Hermitian parts with the Hermitian matrices $\hat K$ and $\hat G^{\text{LS}}$:
\begin{equation}
    \hat J = \delta^2\sum_{\tau,\tau'=-M}^{M}\text{tr}_B\Big(\mathcal{T}(\hat{V}_I(\tau)\hat{V}_I(\tau'))\hat\phi\Big) \equiv \hat K + i\hat G^{\text{LS}}.
\end{equation}
The new matrix $\hat K$ is given by 
\begin{gather}
    \hat K = \frac{\delta^2}{2}\sum_\mu\sum_{\tau = -M}^{M} \sum_{\tau'={-M}}^{\tau-1}\Big( e^{ih\delta(\tau'-\tau)}f(\tau) f(\tau') \hat A_{\mu,I}(\tau)\hat A_{\mu,I}(\tau')+e^{ih\delta(\tau-\tau')}f(\tau) f(\tau') \hat A_{\mu,I}(\tau')\hat A_{\mu,I}(\tau)\Big)  \nonumber \\
    + \frac{\delta^2}{2}\sum_\mu\sum_{\tau=-M}^{M} f(\tau)^2(\hat A_{\mu,I}(\tau))^2 
    \nonumber \\
    = \frac{\delta^2}{2}\sum_\mu\bigg( \sum_{\tau=-M}^{M} f(\tau)^2(\hat A_{\mu,I}(\tau))^2 + \bigg(\sum_{\tau = -M}^{M} \sum_{\tau'={-M}}^{\tau-1}+\sum_{\tau = -M}^{M} \sum_{\tau'={\tau}+1}^{M}\bigg) e^{ih\delta(\tau'-\tau)}f(\tau) f(\tau') \hat A_{\mu,I}(\tau)\hat A_{\mu,I}(\tau')\bigg) \nonumber \\
    = \frac{\delta^2}{2}\sum_\mu\sum_{\tau',\tau=-M}^{M}e^{ih\delta(\tau'-\tau)}f(\tau) f(\tau') \hat A_{\mu,I}(\tau)\hat A_{\mu,I}(\tau') \nonumber \\
    = \frac{1}{2}\sum_\mu \hat L^\dagger_\mu \hat L_\mu.
    \label{eq:Kderivation}
\end{gather}
In going from the first to the second line, we performed a change of variable $\tau \leftrightarrow \tau'$, and used $\sum_{\tau} \sum_{\tau' < \tau} = \sum_{\tau'}\sum_{\tau>\tau'}$. 

The matrix $\hat G^{\text{LS}}$ is given by 
\begin{gather}
    \hat G^{\text{LS}} = \frac{\delta^2}{2i}\sum_\mu \sum_{\tau = -M}^{M} \sum_{\tau'={-M}}^{\tau-1}\Big( e^{ih\delta(\tau'-\tau)}f(\tau) f(\tau') \hat A_{\mu,I}(\tau)\hat A_{\mu,I}(\tau')-e^{ih\delta(\tau-\tau')}f(\tau) f(\tau') \hat A_{\mu,I}(\tau')\hat A_{\mu,I}(\tau)\Big)  \nonumber \\ 
    = \frac{\delta^2}{2i}\bigg(\sum_{\tau = -M}^{M} \sum_{\tau'={-M}}^{\tau-1}-\sum_{\tau = -M}^{M} \sum_{\tau'={\tau}+1}^{M}\bigg) e^{ih\delta(\tau'-\tau)}f(\tau) f(\tau') \hat A_{\mu,I}(\tau)\hat A_{\mu,I}(\tau') \nonumber \\ 
    = \frac{\delta^2}{2i}\sum_{\tau,\tau' = -M}^{M} \text{sgn}(\tau-\tau')e^{ih\delta(\tau'-\tau)}f(\tau) f(\tau') \hat A_{\mu,I}(\tau)\hat A_{\mu,I}(\tau'),
    \label{eq:GLSderivation}
\end{gather}
where the sign function is defined with $\text{sgn}(0)=0$. The anti-time-ordered term is given by 
\begin{equation}
    \hat J^\dagger = \delta^2\sum_{\tau,\tau'=-M}^{M}\text{tr}_B\Big(\mathcal{T}^*(\hat{V}(\tau)\hat{V}(\tau'))\hat\phi\Big) \equiv \hat K - i\hat G^{\text{LS}}.
\end{equation}

Combining Eqs.~(\ref{eq:Lderivation}, \ref{eq:Kderivation}, \ref{eq:GLSderivation}) we arrive at the Lindblad form given in the main text, Eq.~(\ref{eq:lindblad}):
\begin{equation}
        \frac{\mathcal{E}_I(\hat\rho)-\hat{\rho}}{\theta^2} = -i[\hat G^{\text{LS}}, \hat\rho] - \{\hat K,\hat\rho\}+\sum_\mu \hat L_\mu \hat\rho\hat L_\mu^\dagger.
\end{equation}
It is interesting to note that our Lindblad equation has the same form as the so-called `Universal Lindblad Equation', derived in an attempt to restore complete positivity to the Redfield equation \cite{nathan2020universal}. In that case, the `filter function' in the frequency domain enters as the square root of the bath spectral function, $\bar f(\omega) = \sqrt{J(\omega)}$. We leave a detailed investigation of the equivalence between the two to future work. 

\section{Approximations in replacing jump operators with operator Fourier transforms}\label{app:fourier}

The jump operators in Eq.~(\ref{eq:Lderivation}) were derived for the digital channel with a finite reset time $M$. For analytical purposes it is convenient to replace this expression with the continuous-time operator Fourier transform, 
\begin{equation}\label{eq:jump_continuous2}
    \hat L_\mu = \int_{-\infty}^\infty dt\ f^c(t) e^{iht} \hat A^c_{\mu,I}(t),
\end{equation}
and similarly for the Lamb-shift 
\begin{equation}\label{eq:Lambshift_continuous2}
    \hat G^{\text{LS}} = \frac{1}{2i}\sum_\mu \iint_{-\infty}^\infty dtdt'\ \text{sgn}(t-t')f^c(t)f^c(t') e^{ih(t'-t)} \hat A^{c\dagger}_{\mu,I}(t)\hat A^c_{\mu,I}(t').
\end{equation}
Here we use 
\begin{equation}
    f^c(t) \propto \exp \Big(-\frac{a^2t^2}{2}\Big),
\end{equation}
with normalisation $\int_{-T}^Tdt|f^c(t)| = 1$, and
\begin{equation}
    \hat A^c_{\mu,I}(t) = e^{it\hat H_S} \hat A_{\mu,I} e^{-it\hat H_S}
\end{equation}
to denote the continuous time versions of the filter function and interaction picture operators, although we suppress this notation in the main text. Below we consider a single jump operator for simplicity and drop the dependence on $\mu$.

The error in this substitution is controlled by the smallness of the two parameters, $\xi_T = 1/aT$, $\xi_\delta = a\delta$, as we show here. Let 
\begin{equation}
    \hat L(M, \delta) = \delta\sum_{\tau = -M}^{M} f(\tau) e^{ih\delta\tau} \hat A_I({\tau}).
\end{equation}
First we show the bound 
\begin{equation}
    \| \hat L(M,\delta) - \hat L(\infty,\delta)\| \leq 2(aN)^{-1}\|\hat A\| \frac{e^{-T^2a^2/2}}{aT},
\end{equation}
with $N$ the filter function normalisation ($N\sim a^{-1}$) and $\|\hat A\|$ the operator norm of $\hat A$:
\begin{gather}
\| \hat L(M,\delta) - \hat L(\infty,\delta)\| = 
    \Big\| \delta \sum_{\tau={M+1}}^\infty (f(\tau) e^{ih\delta\tau}\hat A_I(\tau) +f(-\tau) e^{-ih\delta\tau}\hat A_I(-\tau))\Big\| \nonumber \\
    \leq 2N^{-1}\|\hat A\| \Big|\delta \sum_{\tau=M+1}^\infty e^{-\delta^2a^2\tau^2/2}\Big | \nonumber \\
    \leq 2N^{-1}\|\hat A\|\int_T^\infty dt \ e^{-a^2 t^2/2} = \sqrt{2\pi} (aN)^{-1} \|\hat A\| \text{erfc}\bigg(\frac{aT}{\sqrt{2}}\bigg),
\end{gather}
where $T=\delta M$, $\text{erfc(x)}$ is the complementary error function, and we used the fact that the Gaussian is monotonically decreasing for $\tau>0$, hence the sum can be bounded by the integral. The error function satisfies the known bound 
\begin{equation}
    \text{erfc}(x) \leq \frac{e^{-x^2}}{x\sqrt{\pi}},
\end{equation}
leading to 
\begin{equation}
    \| \hat L(M,\delta) - \hat L(\infty,\delta)\| \leq 2(aN)^{-1}\|\hat A\| \frac{e^{-T^2a^2/2}}{aT}.
\end{equation}

Next, we consider the frequency domain decomposition of $\hat L$,
\begin{equation}
    \hat L(\infty,\delta) = \delta\sum_{\tau = -\infty}^{\infty} f(\tau) e^{ih\delta\tau} \hat A_I(\tau) = \sum_\omega \hat A_\omega \bigg(\delta \sum_{\tau=\infty}^\infty f(\tau) e^{i(h-\omega)\delta\tau}\bigg). 
\end{equation}
Using the Poisson summation formula \cite{ding2024single} we can replace the infinite time sum with a sum over shifted copies of the continuous-time Fourier transform of the filter function:
\begin{equation}
    \delta \sum_{\tau=\infty}^\infty f(\tau) e^{-i\omega\delta\tau} = \sum_{k=-\infty}^\infty \bar{f}(\omega - 2\pi k/\delta),
\end{equation}
where $\bar f(\omega) = e^{-\omega^2/2a^2}$. Restricting to the range $\omega \delta \ll \pi$, which defines the prethermal regime we work in, the error from keeping only the $k=0$ term in the above sum is of the order $\mathcal{O}(e^{-\pi^2/2\xi_\delta^2})$. Combining the two approximations above, we can replace $\hat L(M,\delta)$ with $\hat L(\infty,0)$ (Eq.~(\ref{eq:jump_continuous2})). 
A similar analysis may be performed for the Lamb-shift term. This gives the final integral-form jump operators appearing in Eqs.~(\ref{eq:jump_continuous2},\ref{eq:Lambshift_continuous2}), and the error involved in these replacements in exponentially small in the parameters $\xi_T$ and $\xi_\delta$.

\section{Explicit form for the Lamb-shift}\label{app:Lamb}

Here we derive an explicit formula for the Lamb-shift Hamiltonian, $\hat G^{\text{LS}}$, which can be expressed in terms of the error-function. To simplify notation we consider a single auxiliary spin, $\hat A_\mu \equiv \hat A$, and work with the continuous time formula given in Eq.~(\ref{eq:Lambshift_continuous2}). 

We first decompose $\hat A = \sum_\omega \hat A_{\omega}$ in terms of cooling transitions of frequency $\omega$, such that $\hat A_I(t) = \sum_\omega \hat A_{\omega}e^{-i\omega t}$. The Lamb-shift is 
\begin{equation}
    \hat G^{\text{LS}} = \sum_{\omega,\omega'} \hat A^\dagger_\omega \hat A_{\omega'} K^{\text{LS}}(\omega,\omega'),
\end{equation}
with the kernel
\begin{gather}
    K^{\text{LS}}(\omega,\omega') = \frac{1}{2i}\iint _{-\infty}^\infty dt dt'\ \text{sgn}(t-t') e^{i[(\omega-h)t-(\omega'-h)t']}f(t)f(t'). 
\end{gather}
Using the Fourier transform of $\text{sgn}(t-t')$, we can write this as a standard principal value integral 

\begin{gather}
    K^{\text{LS}}(\omega,\omega') = \mathcal{P}\int \frac{d\nu}{2\pi \nu} \bar f_h(\omega-\nu) \bar f_h(\omega'-\nu).
\end{gather}
For our choice of Gaussian filter function, $\bar f$ restricts $\omega $ to the range $\omega \approx h+\nu \pm \mathcal{O}(a)$ within the integral; in turn, this restricts the kernel to be finite only when $|\omega-\omega'| \lesssim \mathcal{O}(a)$. To see this explicitly, we may factor
\begin{gather}
    K^{\text{LS}}(\omega,\omega') = \frac{1}{2\pi}\int \frac{d\nu}{\nu-h} e^{-(\omega - \nu)^2/2a^2}e^{-(\omega' - \nu)^2/2a^2} \nonumber \\ 
    = \frac{e^{-(\omega-\omega')^2/4a^2}}{2\pi} \int \frac{d\nu}{\nu+\bar{\omega}-h}e^{-\nu^2/a^2},
\end{gather}
where we defined $\bar{\omega} = \frac{\omega+\omega'}{2}$. The latter integral is the Hilbert transform of a Gaussian, which has a known solution:
\begin{equation}\label{eq:LSkernel}
    K^{\text{LS}}(\omega,\omega') = \frac{1}{\sqrt{\pi}} e^{-(\omega-\omega')^2/4a^2} F\bigg(\frac{\bar{\omega}-h}{a}\bigg).
\end{equation}
The function
\begin{equation}
    F(x) = \frac{\sqrt{\pi}}{2i} e^{-x^2} \text{erf}(ix),
\end{equation}
is a real-valued antisymmetric function known as Dawson's function, and erf is the error function. The main points to note are that $F(x) \approx \frac{1}{2x}$ for large $x$, and $F(x) \approx x$ for small $x$. In total, Eq.~(\ref{eq:LSkernel}) shows that the Lamb-shift kernel is a real-valued function which is symmetric in the argument $\omega-\omega'$, and suppressed as a Gaussian of width $2a$ for large $|\omega-\omega'|$.   
This can be compared to the kernel of the Hamiltonian required for exact detailed balance, derived in Eq.~(\ref{eq:trueLamb}), 
\begin{gather}
    K^{\text{DB}}(\omega,\omega') 
    = i \tanh\bigg(\frac{\beta(\omega-\omega')}{4}\bigg) e^{-(\omega-\omega')^2/4a^2} e^{-(\bar{\omega}-h)^2/a^2}.
\end{gather} 
In contrast to (\ref{eq:LSkernel}), $K^{\text{DB}}$ is an imaginary function, antisymmetric in $\omega - \omega'$, but with the same Gaussian decay for large values of the argument $\omega-\omega'$.

\section{Collected bounds}\label{app:bounds}

In this Appendix, we assemble a collection of bounds that are used in the main text. 

\subsection{Operator norm bounds}
First we bound the operator norms $\|\hat X\|$ (largest singular value of $\hat X$) of the jump operators $\hat L_\mu$ and the Lamb shift $\hat G^{\rm LS}$. We assume throughout the continuous time limits used in Section~\ref{sec:QDB} of the main text.

\subsubsection{Bound on jump operator norm, $\|\hat L_\mu \|$}

The norm on the jump operators as defined in Eq.~(\ref{eq:jump_continuous}) can be bounded as $\|\hat L_\mu\| \leq \|\hat A_\mu \|$, following
\begin{equation}\label{eq:jump_op_bound}
    \|\hat L_\mu\| = \bigg\|\int_{-\infty}^\infty dt\ f(t) e^{iht} \tilde A_\mu(t)\bigg\| \leq \|\hat A_\mu\|\int_{-\infty}^\infty dt\ |f(t)|  = \|\hat A_\mu\|,
\end{equation}
using our normalisation assumption $\int_{-\infty}^\infty dt\ |f(t)|  = 1$ and the unitary invariance of the operator norm. 

\subsubsection{Bound on Lamb-shift operator norm, $\|\hat G^{\rm LS} \|$}

The norm of the Lamb shift defined in Eq.~(\ref{eq:Lambshift_continuous2}) can be bounded as $\|\hat G^{\rm LS}\| \leq \frac{1}{2}\sum_\mu \|\hat A_\mu \|^2 $, following
\begin{gather}\label{eq:LS_op_bound}
    \|\hat G^{\rm LS}\| = \frac{1}{2}\bigg\|\sum_\mu \iint_{-\infty}^\infty dtdt'\ \text{sgn}(t-t')e^{-ih(t-t')}f^*(t)f(t') \tilde A^\dagger_{\mu}(t)\tilde A_{\mu}(t')\bigg\| \nonumber\\
    \leq \frac{1}{2}\sum_\mu \|\hat A_\mu \|^2\iint_{-\infty}^\infty dtdt'\ |f(t)||f(t')|   = \frac{1}{2}\sum_\mu \|\hat A_\mu \|^2.
\end{gather}

\subsection{Bound on induced trace norm of the Lindbladian, $\|\mathcal{L}\|_{1\leftrightarrow 1}$}

Combining these two bounds, we bound the induced trace norm of the Lindbladian $\mathcal{L}$. The induced trace norm of a superoperator $\mathcal{M}$ is defined as 
\begin{equation}
    \|\mathcal{M}\|_{1\leftrightarrow 1} = \sup_{\| \hat X\|_1\leq 1} \|\mathcal{M}(\hat X)\|_1,
\end{equation}
where $\|\hat X\|_1 = \text{Tr}(\hat X^\dagger\hat X)^{\frac{1}{2}}$ denotes the trace norm of the operator $\hat X$. 

The Lindbladian (in the interaction picture) is given by 
\begin{equation}
    \mathcal{L}(\bullet) = -i[\hat G^{\rm LS}, \bullet] + \sum_\mu \Big( \hat L_\mu \bullet\hat L^\dagger_\mu - \frac{1}{2}\{L^\dagger_\mu L_\mu, \bullet\}\Big).
\end{equation}
For the Hamiltonian, $\|[\hat G^{\rm LS}, \hat X]\|_1 \leq 2\|\hat G^{\rm LS}\|\|\hat X\|_1$, and similarly for the dissipative part, $\| \hat L_\mu \hat X\hat L^\dagger_\mu \|_1 \leq \|\hat L_\mu \|^2 \|\hat X \|_1$, $\| \hat L^\dagger_\mu \hat L_\mu \hat X \|_1 \leq \|\hat L_\mu \|^2 \|\hat X\|_1$. Using (\ref{eq:jump_op_bound}) and (\ref{eq:LS_op_bound}), we have 
\begin{equation}
    \| \mathcal{L}\|_{1\leftrightarrow 1} \leq 2 \|\hat G^{\rm LS}\| + 2\sum_\mu \|\hat L_\mu\|^2 \leq 3 \sum_\mu \|\hat A_\mu \|^2 \leq 3N_B\ \sup_\mu \|\hat A_\mu\|^2. 
\end{equation}

\subsection{Bound on the matrix elements of $\Delta \hat G$}

We bound the absolute value of the individual matrix elements of the correction term $\Delta \hat G = \hat G^{\rm LS} - \hat G^{\rm DB}$, in the eigenbasis of $\hat H_S$. From Section \ref{app:Lamb}, $\Delta\hat G$ can be written in the frequency representation as 
\begin{equation}
    \Delta\hat G = 2\sum_{\mu,\omega,\omega'}\hat A^\dagger_{\mu,\omega}\hat A_{\mu,\omega'} e^{-(\omega-\omega')^2/4a^2}y(\omega,\omega'),
\end{equation}
where we defined
\begin{equation}
    y(\omega,\omega') =\frac{1}{2\sqrt{\pi}}F\bigg(\frac{\bar{\omega}-h}{a}\bigg)-\frac{i}{2}\tanh\bigg(\frac{\beta(\omega-\omega')}{4}\bigg)e^{-(\bar{\omega}-h)^2/a^2}.
\end{equation}
This function satisfies $|y(\omega,\omega')|\leq 1$. Thus the matrix elements in the eigenbasis of $\hat H_S$ are bounded as:
\begin{gather}\label{eq:DeltaGbound}
    |(\Delta\hat G)_{ab}| = 2 e^{-\omega_{ab}^2/4a^2} \bigg|\sum_{\mu, \gamma} \langle a |\hat A_\mu^\dagger|\gamma\rangle\langle \gamma|\hat A_\mu|b\rangle y(\omega_{\gamma a}, \omega_{\gamma b})\bigg| 
    \leq 2 e^{-\omega_{ab}^2/4a^2} \sum_{\mu,\gamma} |\langle a |\hat A_\mu^\dagger|\gamma\rangle||\langle\gamma|\hat A_\mu|b\rangle| \nonumber \\
    \leq 2e^{-\omega_{ab}^2/4a^2} \sum_\mu |\bra{a}\hat A^\dagger_\mu \hat A_\mu\ket{a}|^{\frac{1}{2}}|\bra{b}\hat A^\dagger_\mu \hat A_\mu\ket{b}|^{\frac{1}{2}} \nonumber \\
    \leq 2 e^{-\omega_{ab}^2/4a^2}N_B\ \sup_\mu \|\hat A_\mu\|^2.
\end{gather}
In the third line we used the Cauchy-Schwarz inequality.  

\subsection{Bound on steady state coherences}

We now bound the size of individual coherences of the steady state, in the eigenbasis of $\hat H_S$. In the main text, the coherences were derived from the perturbation theory about the Gibbs state, Eq.~(\ref{eq:corrections}), reproduced below:  
\begin{equation}
    \zeta^{\rm coh}_{a\neq b} = i\theta^2(\Delta G)_{ab}\bigg(\frac{p(\epsilon_b) - p(\epsilon_a)} {1 - e^{2iT\omega_{ab}}-i\omega_{ab}\lambda T}\bigg).
\end{equation}
We denote the denominator by $D(\omega_{ab}T) = 1-e^{2iT\omega_{ab}}-i\omega_{ab}\lambda T$, with $T$ the reset time and $\lambda$ the randomization parameter. We first lower bound the denominator. Writing 
\begin{gather}
|D(x)| = |1-\cos2x-i\sin2x-i\lambda x| = \sqrt{(1-\cos 2x)^2+(\lambda x+\sin 2x)^2},
\end{gather}
we then have 
\begin{equation}
    |D(x)| \geq |\lambda x+\sin 2x| \geq \lambda |x|-|\sin 2x | \geq (\lambda -2)|x|.
\end{equation}
Here we used $(1-\cos 2x)^2\geq 0$, the triangle inequality, and $|\sin 2x| \leq |2x|$. The numerator $p(\epsilon_a)-p(\epsilon_b)$ can be upper bounded by rewriting the difference of the Gibbs weights as:
\begin{equation}
    p(\epsilon_a)-p(\epsilon_b) = -(p(\epsilon_a)+p(\epsilon_b)) \tanh \Big (\frac{\beta \omega_{ab}}{2} \Big). 
\end{equation}
A useful upper bound is 
\begin{equation}\label{eq:num_bound_1}
    |p(\epsilon_a)-p(\epsilon_b)| \leq (p(\epsilon_a)+p(\epsilon_b)) \min \bigg(\frac{\beta |\omega_{ab}|}{2}, 1\bigg),
\end{equation}
using $|\tanh x| \leq \min (x,1)$. Combining this bound with the bound on the denominator and Eq.~\ref{eq:DeltaGbound}, we finally have 
\begin{equation}\label{eq:coherences_bound_app}
    |\zeta^{\rm coh}_{a\neq b}| \leq C  \theta^2N_B\  e^{-\omega_{ab} ^2/4a^2} \ \min \bigg[ \frac{\beta}{2T}, \frac{1}{|\omega_{ab}|T}\bigg] \max (p(\epsilon_a), p(\epsilon_b)) ,
\end{equation}
with $C = \frac{1}{\lambda-2} \sup_\mu \|\hat A_\mu\|^2$ an $\mathcal{O}(1)$ constant, and we assue $\lambda > 2$. 

\subsubsection{High temperature bound} 

At high temperatures, we can take the bound independently of $\omega$ and obtain, assuming $\epsilon_b > \epsilon_a$ for simplicity:
\begin{equation}
    |\zeta^{\rm coh}_{a\neq b}| \leq C' \theta^2\beta  N_B  \frac{e^{-\beta \epsilon_a}}{TZ}.
\end{equation}

\subsubsection{Low temperature bound} 

We focus on the coherence between the ground state and low-lying excited states, in the limit $\beta \Delta \gg 1$. The gap is denoted $\Delta$ and we have $|\omega_{0b}|\geq \Delta$. The bound (\ref{eq:coherences_bound_app}) simplifies to 
\begin{equation}
    |\zeta_{0b}^{\rm coh}| \leq C\theta^2N_B\ \frac{e^{-\Delta^2/4a^2} }{\Delta T},
\end{equation}
where we used $\max (p(\epsilon_0),p(\epsilon_b)) = p(\epsilon_0) \leq 1$.

\section{Hamiltonian renormalisation}\label{app:renormalisation}

The corrections to the Gibbs state derived in Section \ref{sec:thermal} were derived using a perturbation theory around the Gibbs state in the Schrödinger picture. In this appendix, we show how the same formulas can be obtained through an alternative method, guided by the following intuition: on physical grounds, we expect that the presence of the bath modifies (`renormalises') the bare system Hamiltonian (leading to e.g.~level shift phenomena known from atomic theory \cite{cohen2024photons}), hence there is no reason to expect exact thermalization into the Gibbs state set by $\hat H_S$ \cite{trushechkin2022open}. While detailed balance is therefore broken in the interaction picture with respect to the \emph{bare} Hamiltonian, it may be that it is restored in the interaction frame with respect to a \emph{renormalised} Hamiltonian, $\hat H'_S$.  We show below that this intuition is correct, and that the renormalised system Hamiltonian can be solved for self-consistently at second order in the coupling $\theta$. The solution does not fix the diagonal part of the Hamiltonian, which controls the populations. 

We define the renormalised Hamiltonian with the correction expanded in powers of the system-bath coupling
\begin{equation}\label{eq:renormalisedH}
    \hat H'_S = \hat H_S + \theta^2 \hat C, \hspace{1cm} \hat C = \sum_{k=0}^\infty \theta^{2k} \hat C_k,
 \end{equation}
 where $\hat C$ is an operator acting only on the system Hilbert space. We assume throughout that the Trotter angle is small such that 
 \begin{equation}
     e^{-i\delta\theta^2\hat C} e^{-i\delta \hat H_S} \approx e^{-i\delta\hat H'_S} \equiv\hat  U_S'.
 \end{equation}
Then, by making use of the identities 
\begin{equation}
 = e^{-i\delta\theta\hat V(\tau)} e^{i\delta\theta^2\hat C}e^{-i\delta \hat H_B} e^{-i\delta\theta^2\hat C}e^{-i\delta\hat H_S} = e^{-i\delta\theta \hat V(\tau)}e^{i\delta\theta^2\hat C}\hat U_B \hat U'_S,
\end{equation}
\begin{equation}
    \hat R = \prod_{\tau = 1}^{M_R} e^{-i\delta\tau \hat H_S} = \prod_{\tau = 1}^{M_R} e^{i\delta \theta^2\hat C}\hat U'_S
\end{equation}
we define the interaction picture with respect to the renormalised Hamiltonian as 
\begin{equation}
    \hat{O}_{I_R}(\tau) = \hat Y_0^{-\tau} \hat O(\tau)\hat Y_0^\tau,, \hspace{.5cm} \hat Y_0 = \hat U_B \hat U'_S,
\end{equation}
and $\hat{Q}_{I_R} = \hat U^{\prime - M_R}_S\hat Y_0^{-M} \hat Q \hat Y_0^{-(M+1)}$. The Schrödinger map is then written in the form 
\begin{gather}
    \mathcal{E}(\hat \rho) = 
    \mathcal{S'}^{M+M_R} \circ{\mathcal{E}}_{I_R}\circ\mathcal{S'}^{M+1}(\hat\rho),
\end{gather}
\begin{equation}
    \mathcal{S'}(\hat\rho) = \hat U'_S(\hat\rho)\hat U^{\prime-1}_S, \hspace{1cm}
   {\mathcal{E}}_{I_R}(\rho) = \text{tr}_B\big[ {\hat Q_{I_R}}(\hat \rho\otimes \hat \phi)\hat Q_{I_R}^{\dagger}\big].
\end{equation}

The weak-coupling derivation follows the same steps as laid out in Appendix \ref{app:expansion}. The only difference at leading order in $\theta$ is the appearance of an additional coherent term:
\begin{equation}\label{eq:lindblad_rn}
    \frac{{\mathcal{E}}_{I_R}(\hat\rho)-\hat{\rho}}{\theta^2} = -i[\hat G^{\text{LS}}-\hat G^C, \hat\rho]+ \{\hat K,\hat\rho\}+\sum_\mu \hat L_\mu \hat\rho\hat L_\mu^\dagger,
\end{equation}
where 
\begin{equation}
    \hat G^C = \delta \sum_{\tau=-M}^{M+M_R}\hat {C}_{0,I_R}(\tau) \approx \int_{-T}^{T+T_R} dt\  \hat{C}_{0,I_R}(t).
\end{equation}

We next seek the frame in which the quantum detailed balance equations are satisfied. Denote by $\phi_a'$ the basis of eigenstates of $H'_S$. The filter function guarantees that Eq.~(\ref{eq:jump_transition1}) holds with the replacement $\phi_a \to \phi'_a$, and hence the dissipative part of Eq.~(\ref{eq:lindblad_rn}) still satisfies QDB. To satisfy QDB for the coherent terms, we require that 
\begin{equation}
    \hat G^{\text{LS}}-\hat G^{\text{DB}}-\hat G^C + \hat \Lambda = 0,
\end{equation}
where $\hat \Lambda$ is a term diagonal in the eigenbasis of $\hat H_S'$. If we can find $\hat C$ such that the above condition is satisfied, then the steady state of the channel ${\mathcal{E}}_{I_R}$ is given by the QDB solution 
\begin{equation}\hat \sigma'_\beta \propto e^{-\beta \hat H'_S}.
\end{equation}
Returning to the Schrödinger picture, $\hat Y_0$ commutes with $\hat \sigma'_\beta$, and therefore $\hat \sigma'_\beta$ is also the \emph{exact} steady state for the cooling channel, $\mathcal{E}(\hat\sigma'_\beta) = \hat\sigma'_\beta$. 

The solution which satisifes QDB is simply found in the basis of eigenstates of $\hat H'_S$. Ignoring for a moment the randomization step and setting $T_R=0$, we have 
\begin{equation}
    (\hat C_0)_{a'b'} = \frac{(\Delta \hat G)_{a'b'}}{\int_{-T}^T dt \ e^{-i\omega'_{ab}t}}+\delta_{ab}\frac{\Lambda_{a}}{2T} = \frac{(\Delta \hat G)_{a'b'}}{2T\text{sinc}(\omega'_{ab}T)} +\delta_{ab}\frac{\Lambda_{a}}{2T},
\end{equation}
where the primes remind us that we are working in the eigenbasis of $\hat H'_S$. Returning to the eigenbasis of $\hat H_S$ requires a non-trivial perturbation expansion. However, since we only require $\hat C_0$ to leading order, we can substitute the original eigenbasis in place above $\phi'_a \to\phi_a$ i.e.~dropping the primes (assuming that the perturbation $\hat C$ did not lead to degeneracy splitting). 

The above expression is problematic due to the resonances at $\omega_{ab}T = k\pi$, as we observed in Section \ref{sec:thermal}. To include the randomization step, we should consider the channel averaged over the random time $T_R$. Note that now both the interaction-picture map and the Schrödinger-picture map depend on $T_R$. The correct average is performed by fixing
\begin{equation}
    \mathbb{E}\ \int_{-T}^{T+T_R} dt\ (\hat U'_S)^{T_R}\hat{C}_{0,I_R}(t)(\hat U_S^{\prime -1})^{T_R} = \mathbb{E}\ (\hat U'_S)^{T_R}
    \Delta\hat G(\hat U_S^{\prime -1})^{T_R},
\end{equation}
which after a little algebra gives
\begin{equation}
    (\hat C_0)_{ab} = \frac{i\omega_{ab}e^{i\omega_{ab}T}}{1-i\omega_{ab}T_0-e^{2i\omega_{ab}T}}(\Delta \hat G)_{ab}+\delta_{ab}\frac{\Lambda_a}{2T+T_0}.
\end{equation}
It is unfortunate that the diagonal correction $\hat \Lambda$ remains undefined in this scheme. However, we expect that our approach of solving for the renormalised Hamiltonian by moving to the self-consistent frame satisfying detailed balance could be of use more generally, for example in determining the Hamiltonian of `mean force' which arises from a finite coupling to a macroscopic bath \cite{trushechkin2022open, lee2022perturbative, correa2023potential, timofeev2022hamiltonian}.

\end{widetext}

\bibliography{bib}
\end{document}